\documentstyle[12pt]{article}

\begin{document}

\thispagestyle{empty}
\begin{center}
{\Large\sc Non-Riemannian Gravitation and its}
\\[5pt]
{\Large\sc relation with Levi-Civita theories}
\\
\ \\
by
\\
{\bf Roberto Scipioni}
\\
(Honour Degree University of Rome)
\\
\ \\
\begin{sl}
School of Physics and Chemistry,
Lancaster University,
England
\end{sl}
\\
\ \\
Thesis submitted for the degree of Doctor of Philosophy
\\
Lancaster University,
\\
December 1998 
\\
{\ }
\\
{\ }
\\
{\ }
\\
{\ }
\\
{\Large\em Dedicated to my family}
\end{center}
\newpage
\thispagestyle{empty}
\begin{center}
{\bf Original Contribution}
\end{center}
\bigskip
3.3\\
Chapter 4\\
Chapter 5\\
6.2, 6.3\\
Chapter 7\\
8.2, 8.4\\
\newpage
\thispagestyle{empty}
\begin{center}
{\bf ACKNOWLEDGEMENTS}
\end{center}
There are many people who I would like to thank. Let me begin with Cardinal Achille and Professor Bruno Silvestrini who trusted me and helped me in obtaining the funding from the International Centre for Cultural Cooperation and Development NOOPOLIS (Italy), which funded my stay in Lancaster.\\
I wish to thank in the first place my friend Emanuele Costa now at Edinburgh University with whom I shared most of my experience in Lancaster, he was not only a friend but also a guide especially during the difficult moments of my permanence in Lancaster. The interaction with many colleagues was invaluable first of all Dr. Charles Wang for teaching me many things about computer usage; his confidence in the frame independent approach to differential geometry was always inspiring for me.\\
Thanks to some of my friends especially for the last period of my stay in Lancaster, Hana Chaudhry, Kaori Sasaki, Emilio Guaglianone. My gratitude goes to my family in Rome for their love and patience and for helping me financially.\\
I could not finish of course without mentioning my supervisor Prof. Robin Tucker, to whom I owe everything I know about non-Riemannian gravity and for being kind during the difficult moments of my stay in Lancaster.
\newpage
\begin{center}
{\bf CONTENTS}
\end{center}
\bigskip
{\bf 1. INTRODUCTION}\\
\\
{\bf 2. NON-RIEMANNIAN GEOMETRY}\\
\\
{\bf 3. \bf TENSORIAL VARIATIONAL METHODS}\\
\\
...3.1 \emph{Divergence conditions}\\
\\
...3.2 \emph{Variations of the generalised Eintein-Hilbert action}\\
\\
...3.3 \emph{Particular solutions of the Cartan equation}\\
\\
{\bf 4. NON-RIEMANNIAN GRAVITATION AND THE ISOMORPHISM WITH GENERAL RELATIVITY}\\
\\
...4.1 \emph{Relation with Einstein-Proca theory: massless Weyl field}\\
\\
...4.2 \emph{Massive Weyl field}\\
\\
...4.3 \emph{Extended actions}\\
\\
...4.4 \emph{Conformal generalised Einstein equations}\\
\\
...4.5 \emph{Relation with a class of Dilaton gravity theories}\\
\\
{\bf 5. EXACT SOLUTIONS OF NON-RIEMANNIAN GRAVITY}\\
\\
...5.1 \emph{The axially symmetric static Melvin solution}\\
\\
...5.2 \emph{The time dependent Rosen solution}\\
\\
...5.3 \emph{Non-Riemannian black-hole with Dilaton}\\
\\
...5.4 \emph{Black-hole solutions with more general actions}\\
\\
{\bf 6. MODIFICATION OF THE MAXWELL THEORY IN NON-RIEMANNIAN GRAVITY }\\
\\
...6.1 \emph{Conservation axioms and Maxwell theory}\\
\\
...6.2 \emph{Field equations with the term} $\star (T^c \wedge e_{c}) \wedge F \wedge A$\\
\\
...6.3 \emph{Field equations with the term} $(F \wedge F)\star(Q \wedge \star Q)$\\
\\
{\bf 7 CONCLUDING REMARKS}\\
\\
\newpage
{\bf 8. APPENDICES}\\
\\
{\bf No-hair theorem for spherical scalar black hole}\\
\\
...8.1 \emph{Riemannian case}\\
\\
...8.2 \emph{Non-Riemannian case}\\
\\
{\bf Autoparallels in non-Riemannian gravity}\\
\\
...8.3 \emph{Autoparallels in non-Riemannian gravity}\\
\\
...8.4 \emph{The 'Proca-type' case}\\
\\
{\bf Differential Geometry and Exterior Calculus}\\
\\
...8.5 \emph{Differential Geometry and exterior calculus}\\
\newpage
\begin{center}
{\bf ABSTRACT}
\end{center}
\bigskip
\bigskip
\bigskip
In this thesis the relation between certain models of non-Riemannian gravity and Levi-Civita theories are investigated; in particular the simplification which occurs in the field equations which are obtained from a variational principle. This relation was originally discovered by Tucker, Wang and others [25,26] who considered either the case of a 4-dimensional space time or the more general case of a $n$-dimensional spacetime with the help of the program Maple [1] and limited their analysis to the Einstein-Proca case.\\
So far a proof by hand valid for an $n$-dimensional spacetime has not been presented.\\
In this thesis a step by step proof is presented of the remarkable property of simplification in the field equations. To obtain this proof we use a systematic study of some elementary but relevant solutions of the Cartan equation for the non-Riemannian part of the connection. These solutions will be used to calculate the terms which appear in the generalised field equations.\\
The improved insight obtained by this proof is then used to extend the result to other gravity theories like some models with scalar fields. We apply this theorem for obtaining some exact new solutions, in particular a new class of non-Riemannian black-hole solutions with Dilaton fields. We briefly discuss the relation with the no-hair theorem in appendix A, in there we briefly discuss why the no-hair theorem cannot be extended to non-Riemannian gravity.\\
Then we move to another interesting problem in non-Riemannian gravity; the possible modification of Maxwell theory in non-Riemannian gravity. We follow Hehl and others [36] to show that if we start from the axioms of conservation, only the constitutive law has to be modified. Using the results of chapter 2 and 3 we show with two examples how the study of the modified Maxwell theory can be obtained in the non-Riemannian case.\\
In appendix B we obtain the expression of the autoparallels for a certain class of non-Riemannian models of gravitation.
\newpage
\begin{center}
\bf {CHAPTER 1}
\end{center}
\bigskip
\section{INTRODUCTION}
Einstein's theory of gravity which was formulated more than eighty years ago provides an elegant and powerful formulation of gravitation in terms of a pseudo-Riemannian geometry. Einstein's equations are obtained by considering variations with respect to the metric of the integral of the curvature scalar associated with the Levi-Civita connection, with respect to the volume form of the spacetime. Einstein assumed that the connection was metric compatible and torsion free; a position which is natural but not always convenient. 
A number of developments in physics recently have suggested the possibility that the treatment of spacetime might involve more than just a Riemannian spacetime for example:\\
\\
1] The vain effort to quantise gravity is perhaps so far a strong piece of evidence for going beyond a geometry which is dominated  by the classical concept of distance.\\
\\
2] The generalisation of the theory of elastic continua with structure to 4-dimensional spacetime suggests, physical interpretations of the non-Riemannian structures which emerge in the theory [45,46].\\
\\
3] The description of Hadron (or nuclear) matter in terms of extended structures [47,48].\\
\\
4] The study of the early universe, in particular singularity theorems, the problem of the unification of interactions and the related problem of compactification of dimensions, models of inflation with dilaton-induced Weyl covector [49].\\ 
\\
Moreover at the level of the so called string theories there are hints [2-5] that by using non-Riemannian geometry we may accommodate the several degrees of freedom coming from the low energy limit of string interactions in terms of a non metric compatible connection with torsion. It is interesting to observe that since string theories are expected to produce effects which are at least in principle testable at low energies, there may be chances to obtain non-Riemannian models with predictions which can somehow be tested, moreover some models may have some effects on astronomical scales [6-16]. For instance, recently models have been proposed that permit us to account for the so called dark matter by invoking non-Riemannian gravitational interactions [17]. 
There are several approaches to non-Riemannian gravity, perhaps one of the most popular is the one which uses the gauge approach [18-21].\\
Soon after Einstein proposed his gravitational theory, Weyl found an extension to it, to include electromagnetism in a unified way [50]. Weyl's theoretical concept was the so called \emph{gauge invariance of length}. For that purpose, Weyl extended the geometry of spacetime from the Levi-Civita connection to a Weyl space with an additional covector $Q = Q_{a} e^{a}$ where $e^{a}$ denotes the field of coframes of the four-dimensional manifold.\\
The Weyl connection 1-forms reads:
\begin{eqnarray*}
\Gamma^{W}{}_{\alpha \beta} = \Gamma_{\alpha \beta} + \frac{1}{2}(g_{\alpha \beta} Q - e_{\alpha}Q_{\beta} + e_{\beta}Q_{\alpha})
\end{eqnarray*}
The Weyl form is related to the so called non-metricity of the spacetime. If we write the interval in the form:
\begin{eqnarray*}
ds^2 = g_{\mu \nu}dx^{\mu}dx^{\nu}
\end{eqnarray*}
The square length of a generic vector $V$ can be written as:
\begin{eqnarray*}
V^2 = g_{\mu \nu} V^{\mu}V^{\nu} = g(V,V)
\end{eqnarray*}
where $g$ is the (2,0) symmetric tensor defined by:
\begin{eqnarray*}
g = g_{\mu \nu} \, e^{\mu} \otimes e^{\nu}
\end{eqnarray*}
We find that the covariant derivative of $V^2$ with respect to a generic vector $X$ gives:
\begin{eqnarray*}
\nabla_{X} V^2 = (\nabla_{X} g) (V,V) = Q_{ab}(X) V^{a}V^{b}
\end{eqnarray*}
where $Q_{ab}(X) = ({\nabla_{X}}g)(X_{a},X_{b})$, $Q^{a}{}_{a} = Q$ and we supposed that ${\nabla_{X}}V = 0$.\\
If a spacetime has a nonmetricity then if we parallel transport a vector along a curve whose tangent vector is $X$, the length changes.\\
In Weyl's theory the field $Q$ is identified with the electromagnetic potential $A$ [50].\\
Subsequently it was found that Weyl's theory is not viable. However the concept of gauge invariance survived. In particular the concept of local gauge invariance flourished in the field of theoretical particle physics. Suppose we have a wave function describing a particle in quantum mechanics then we may postulate the invariance of the theory under the $U(1)$ invariance group:
\begin{eqnarray*}
\Psi \rightarrow e^{i \alpha(x)} \Psi,
\end{eqnarray*}
with $\alpha(x)$ function of spacetime.\\
Using this approach we can construct the whole classical Dirac-Maxwell theory, indeed we can get the electromagnetic interaction by the requirement of the local invariance under the U(1) gauge group.\\
Yang and Mills, in 1954 generalised the abelian $U(1)$ gauge invariance to non-Abelian $SU(2)$-gauge invariance using the approximately conserved isotopic spin current as starting point.\\
In any case it is interesting to observe that the gauge principle originated from General Relativity.\\
Nowadays the notion of gauge symmetry is one of the cornerstones of theoretical physics, the three non-gravitational interactions are described by means of gauge theories in the framework of the standard model.\\
Thanks to the works of Utiyama, Sciama and Kibble [21,51,52] it was realized that also gravitation can be formulated as a gauge theory. The gauge group being in that case the \emph{Poincare` group} that is the semidirect group of the translations and the Lorentz group.\\
More recently a quite general gauge theory has been formulated which includes General Relativity as a particular case; in the latter case the gauge group is the so called \emph{affine group} resulting from the semidirect product of the translations and the general linear group $GL(n,R)$. This theory is called \emph{Metric Affine Gravity} [18], in general these theories allow for the introduction of a general non-Riemannian connection.\\
Though it is possible to treat gravity using the gauge approach, it is necessary to remember that in the case of gravity contrary to the case of strong and electroweak interactions we are considering an external symmetry group i.e. acting on the spacetime. So a procedure is necessary to mediate the transition from the internal structure which is proper of any gauge formulation and external structures and to project geometric gauge structures on the base manifold to induce gravity. It seems to be unclear how this procedure applied to any affine frames takes place. This issue is somehow similar to the compactification of higher-dimensional supergravity or string theories.\\
 Recently a different approach has been proposed [22-24] based on the metric $ \bf g $ and the connection $ \bf \nabla $ as independent variables. Instead of working with the group $GL(4, \bf R)$ (the general linear group), they rely on the definition of torsion and non-metricity in terms of $ \bf g $ and $ \bf \nabla $.\\
The study of non-Riemannian theories of gravitation is in general quite complicate and a powerful formalism is needed to simplify the calculations. In order to obtain that, the frame independent approach to differential geometry seems quite appropriate.\\
In chapter 2 we introduce the formalism of non-Riemannian geometry using modern differential geometry, starting from the definitions of non-metricity and torsion tensors in terms of $ \bf \nabla $ and $ \bf g $ respectively.\\
In chapter 3 we describe the variational techniques using a tensorial approach. This approach has several advantages, indeed we do not need  to consider particular reference systems or co-frames and the results are unambiguous and easy to apply, bearing in mind that whenever requested, transition to the more traditional component manipulation is possible, taking into account the fact that since the theory is non-metric, raising and lowering of indices must be done with care. In the last paragraph some particular solutions of the Cartan equation are obtained and classified. This classification will be necessary for the following chapters.\\ 
In chapter 4 we investigate the relation of non-Riemannian theories with general relativity. It has been observed few years ago that the field equations for certain non-Riemannian theories of gravitation can be dramatically simplified but so far an explicit proof has not been given especially for the $n$ dimensional case. In several steps we give a proof by hand valid for any $n$ of the relation with Einstein-Proca systems, then we extend the proof to a class of Dilaton gravity theories. The proof is obtained using the systematic study of the particular solutions of the Cartan equation presented in section 3.3. This proof clarifies and generalises what previously found by Tucker, Wang and others [22-26].\\
In chapter 5 we apply this theorem to exhibit some simple but new solutions of non-Riemannian gravity, we consider first the axially symmetric but static Melvin solution, then we consider the time dependent Rosen solution, then we move to the discussion of the possible black-hole solutions in non-Riemannian gravity. By importing results from general Dilaton theories of gravity we are able to exhibit a new class of non-Riemannian black-hole solutions with Dilaton fields. In appendix A we present some material regarding the no-hair theorem with scalar fields. The simple proof of the theorem which uses a conformal mapping between the Penney solution [27] and a general action in scalar theory of gravity is presented.\\
In appendix B we obtain the expression for the autoparallels in non-Riemannian gravity and then we discuss the form they assume for 'Proca-type' theories. In the last chapter we discuss the problem of introducing modifications to the electromagnetic theory. It has been claimed that Maxwell theory may be modified if non-Riemannian connections are present. Following Hehl and others, if we start from the postulates of magnetic and electric charges conservation, only the constitutive law containing the metric may be modified. The conclusion is that the only way to modify the Maxwell theory is to introduce a term in the action dependent on the torsion and non-metricity whose variation gives a contribution to the usual law $d \star F = 0$. Using the results of chapter 2 and 3 we present examples of modifications to the constitutive law. In the first case we add the term $\star(T^{a} \wedge e_{a}) \wedge F \wedge A$ to the action density. In this case we relate the modification of the constitutive law to the torsion of the spacetime. Then we consider the effect of the term $(F \wedge F) \star(Q \wedge \star Q)$, we get the field equations, and we solve for the non-metricity and torsion.\\
\newcommand{\be}{\begin{equation}}
\newcommand{\ee}{\end{equation}}
\newcommand{\z}{\stackrel{-}}
\newpage
\begin{center}
\bf {CHAPTER 2}
\end{center}
\bigskip
\section{NON-RIEMANNIAN GEOMETRY}
\bigskip
In this chapter a brief introduction is given of the non-Riemannian geometry in the frame independent approach.\\ 
The formalism we will be using takes into account the fact in general we will consider non-metric theories, that is theories in which the metric is not regarded any longer as covariantly constant. It is clear that we need to use a formalism which makes use of as less number of indices as possible, this is obtained by formulating the non-Riemannian geometry in the frame independent approach; some important definitions which are used throughout this thesis are collected in Appendix C.\\ 
One of the fundamental concepts in differential geometry is of Parallel Transport. To define parallel transport we need to introduce a linear connection, a type preserving derivation on the algebra of tensors fields commuting with contractions. We will denote such a connection by  $\bf {\nabla}$ . We can specify the most general linear connection by calculating its effects on an arbitrary local basis of vector fields $ {X_{a}} $
\begin{equation}
{\nabla}_{X_{a}} X_{b} = {\Lambda}^{c}{}_{b}(X_{a})X_{c}
\end{equation}
where $ \Lambda^{a}{}_{b} $ are a set of $ n^2$ 1-forms, and $n$ is the dimension of the manifold.\\
It is possible to specify a general connection by giving a (2,0) metric symmetric tensor $ \bf g$, a (2,1) tensor $ \bf T$ defined by:
\be
{\bf T}(X,Y) = {\bf \nabla}_{X}Y - {\bf \nabla}_{Y}X - [X,Y]
\ee
with $X,Y$ vector fields and $\bf S$ a (3,0) tensor symmetric in the last two arguments .
$\bf T$ is the torsion associated with $ \bf \nabla $ and $ \bf S$ is taken to be the metric gradient, $ \bf S = \nabla g $. Then it is possible to calculate the connection as a function of $ \bf g$, $ \bf S $, $\bf T$, indeed by using the relation:
\begin{equation}
X({\bf g}(Y,Z)) = {\bf S}(X,Y,Z) + {\bf g}({\bf {\nabla}}_{X} Y,Z) + {\bf g}(Y,{\bf \nabla}_X Z)
\end{equation}
we obtain
\begin{eqnarray}
2 {\bf g}(Z, {\bf \nabla}_{X}Y) = X({\bf g}(Y,Z)) + Y({\bf g}(Z,X)) - Z({\bf g}(X,Y)) - {\bf g}(X,[Y,Z])\\ \nonumber
- {\bf g}(Y,[X,Z]) - {\bf g}(Z,[Y,X]) - {\bf g}(X,{\bf T}(Y,Z)) -{\bf g}(Y,{\bf T}(X,Z))- \\ \nonumber
{\bf g}(Z,{\bf T}(Y,X)) - {\bf S}(X,Y,Z) - {\bf S}(Y,Z,X) + {\bf S}(Z,X,Y)
\end{eqnarray} 
where $X,Y,Z$ are any vector fields.\\
We define the general curvature operator as:
\begin{equation}
{\bf R}_{X,Y}Z = {\bf \nabla}_{X}{\bf \nabla}_{Y} Z - {\bf \nabla}_{Y}{\bf \nabla}_{X} Z - {\bf \nabla}_{[X,Y]} Z
\end{equation}
which is a type preserving tensor derivation on the algebra of tensor fields. The (3,1) curvature tensor is defined by:
\begin{equation}
{\bf R}(X,Y,Z,\beta) = \beta(R_{X,Y} Z)
\end{equation}
with $\beta$ an arbitrary 1-form. We can introduce the following set of local curvature 2-forms $R^{a}{}_{b}$:
\begin{equation}
R^{a}{}_{b}(X,Y) = \frac{1}{2} {\bf R}(X,Y,X_{b},e^{a})
\end{equation}
where $e^{a}$ is any local basis of 1-forms dual to $X_{c}$. We have $e^{a}(X_{b}) = \delta^{a}{}_{b}$ or by using the contraction operator with respect to $X$, $i_{X_{b}}(e^{a}) = e^{a}(X_{b}) = \delta^{a}{}_{b} $.\\
In terms of the connections forms we can write:\\
\begin{equation}
R^{a}{}_{b} = d {\Lambda}^{a}{}_{b} + {\Lambda}^{a}{}_{c} \wedge {\Lambda}^{c}{}_{b}
\end{equation}
In a similar manner, the torsion tensor gives rise to a set of local 2-forms $T^{a}$:
\begin{equation}
T^{a}(X,Y) \equiv \frac{1}{2}(e^{a}(T(X,Y))
\end{equation}
which can be written:
\begin{equation}
T^{a} = d e^{a} + \Lambda^{a}{}_{b} \wedge e^{b}
\end{equation}
By using the symmetry of the tensor {\bf g}, the tensor {\bf S} can be used to define the local non-metricity 1-forms $Q_{ab}$ symmetric in their indices:
\begin{equation}
Q_{ab}(Z) = {\bf S}(Z,X_{a},X_{b})
\end{equation}
It is convenient very often to make use of the exterior covariant derivative $D$, for its definition see appendix C or Ref. [44].\\ 
With $g_{ab}\equiv {\bf g}(X_{a},X_{b})$ we get that:\\ 
\begin{eqnarray}
Q_{ab} = D g_{ab}\\ \nonumber
Q^{ab} = - Dg^{ab}
\end{eqnarray}
As usual indices are raised and lowered with components of the metric  in a certain local basis. We denote the metric trace of these forms as:
\begin{equation}
Q = Q^{a}{}_{a}
\end{equation}
We call $Q$ the Weyl 1-form.\\
In Riemannian geometry we require that the connection be metric compatible $(Q_{ab}=0$, or equivalently ${\bf S} = 0)$ and ${\bf T} = 0$.\\
A geometry with a torsion-free connection that preserves a conformal metric is known as a Weyl space [28].\\
We can decompose the connection ${\bf \nabla}$ into parts that depend on the Levi-Civita connection $ {\bf \stackrel{o}{\nabla}} $. To this aim we introduce the tensor ${\bf \lambda}$
\begin{equation}
{\bf \lambda} (X,Y,\beta) = \beta({\bf \nabla}_{X} Y) - \beta({\bf \stackrel{o}{\nabla}}_{X}Y)
\end{equation}
for arbitrary vector fields $X,Y$ and 1-form $\beta$.\\
To the decomposition above corresponds a splitting of the connection 1-forms into its Riemannian and non-Riemannian parts ${\Omega}^{a}{}_{b}$ and ${\lambda}^{a}{}_{b}$ respectively as:
\begin{equation}
\Lambda^{a}{}_{b} = \Omega^{a}{}_{b} + \lambda^{a}{}_{b}
\end{equation}
where
\begin{equation}
\lambda^{a}{}_{b} \equiv {\bf \lambda}(-,X_{b},e^{a})
\end{equation}
In terms of these forms we get:
\begin{eqnarray}
T^{a} = \lambda^{a}{}_{c} \wedge e^{c}\\ \nonumber
Q_{ab}= -(\lambda_{ab}+\lambda_{ba})
\end{eqnarray}
Using relation (11) and (17) we get $Q_{ab} = {\bf S}(-, X_{a},X_{b}) = -(\lambda_{ab} + \lambda_{ba})$, then we get:
\be
{\bf S} = 0 \leftrightarrow \lambda_{ab} = - \lambda_{ba}
\ee
That is the metric compatibility requires $\lambda_{ab} = - \lambda_{ba}$.\\
We define the 1-form  $T$ by:
\be
T = i_{a}T^{a}
\ee
With $ i_{a} \equiv i_{X_{a}}$. We have the following relation:
\be
e^c \wedge \star T_{c} = - \star T
\ee
where use has been made of the property $\star(A \wedge e^{a}) = i^{a} \star A$ with $A$ a generic form and $\star$ indicates the Hodge operation associated with the metric ${\bf g}$ (see appendix C).\\
Since a general connection is neither symmetric nor antisymmetric particular care has to be taken when writing the indices because in general $\lambda^{a}{}_{b}$ is different from $\lambda_{b}{}^{a}$.\\
Observe that a Riemannian connection $\Omega^{a}{}_{b}$ being torsion free implies:
\be
de^{a} + \Omega^{a}{}_{b} \wedge e^{b} =0 
\ee
It follows from relation (4) that:
\be
2\Omega_{ab} = (g_{ac}i_{b}-g_{bc}i_{a}+e_{c}i_{a}i_{b})de^{c} + (i_{b}dg_{ac}-i_{a}dg_{bc})e^{c} + dg_{ab}
\ee
and
\be
2\lambda_{ab} = i_{a}T_{b} - i_{b}T_{a} -(i_{a}i_{b}T_{c} + i_{b} Q_{ac} - i_{a}Q_{bc})e^{c}-Q_{ab}
\ee
The field equations of the Einstein theory are obtained as variational equations deduced from the Einstein-Hilbert action, this being the integral of the curvature scalar of the Levi-Civita connection with respect to the volume form.\\
The scalar curvature is obtained by contracting the Ricci-tensor which is the trace of the curvature tensor.\\
In the following we will consider two types of tensors:
\be
{\bf Ric}(X,Y)=e^{a}({\bf R}_{X_{a}X}Y) 
\ee
and
\be
{\bf ric}(X,Y)=e^{a}({\bf R}_{XY}X_{a})
\ee
We have:
\begin{eqnarray}
{\bf Ric}_{cb}= R_{acb}{}^{a}\\ \nonumber
{\bf ric}_{cb}=R_{cba}{}^{a}
\end{eqnarray}
Where:
\be
R_{acb}{}^{d}= {\bf R}(X_{a},X_{c},X_{b},e^{d})
\ee
the first one has no symmetry in general while $\bf ric$ is a two form which can be shown to be:
\be
{\bf ric} = 2R^{a}{}_{a} = - dQ
\ee
Indeed from relation (15,22,23) we get:
\be
2 \Lambda = 2 {\Lambda}^{a}{}_{a} =  -Q
\ee
and from relations (6,7) and (25) we get:
\be
{\bf ric} = 2 R^{a}{}_{a} = 2 d \Lambda = -dQ
\ee
Since $ {\Lambda}^{a}{}_{c} \wedge {\Lambda}^{c}{}_{a} = -{\Lambda}^{c}{}_{a} \wedge {\Lambda}^{a}{}_{c} =  0$.\\
In the Riemannian case $ric = 0$ , and $Ric(X,Y)$ goes into the usual Ricci tensor which is symmetric in the two arguments, this being due to the fact non-metricity is zero, moreover the Riemannian curvature tensor satisfies the antisymmetry property.
\be
R_{abc}{}^{d} = -R_{abd}{}^{c}
\ee
The symmetric part of $Ric$ can be contracted with the metric tensor to obtain a generalised curvature:
\be
R = Ric(X_{a}, X_{b}){\bf g}(X^{a},X^{b}) = Ric(X_{a},X^{a})
\ee
It is possible to obtain the expression for the general $Ric(X,Y)$ as:
\be
Ric(X_{a},X_{b}) = {\stackrel{o}{Ric}}(X_{a},X_{b})+i_{a}i_{c}({\stackrel{o}{D}}\lambda^{c}{}_{b}+ \lambda^{c}{}_{d} \wedge \lambda^{d}{}_{b})
\ee
where $\stackrel{o}{D}$ is the covariant exterior derivative with respect to the Levi Civita connection $\Omega^{a}{}_{b}$ and
\be
R = {\stackrel{o}{R}}+i_{a}i_{c}({\stackrel{o}{D}}\lambda^{ca}+ \lambda^{c}{}_{d} \wedge \lambda^{da})\ee 
If we define the Ricci 1-forms by:
\be
P_{b} = i_{a} R^{a}{}_{b}
\ee
hence
\be
P_{a} = {\bf Ric}(X_{b},X_{a})e^b
\ee
and
\be
R = i^b P_{b} = i^b i_{a} R^{a}{}_{b}
\ee
observe that the general curvature tensor can be written as:
\be
R = 2R^{d}{}_{c} \otimes e^c \otimes X_{d}
\ee 
\newpage
\begin{center}
\bf {CHAPTER 3}
\end{center}
\bigskip
\section{TENSORIAL VARIATIONAL METHODS}
\bigskip
A powerful way to get the field equations of a non-Riemannian theory is to start from a variational principle. Since we are using the frame independent approach, we will use that approach in calculating variations too.\\
In this chapter we consider the equations obtained by local extrema of action functionals defined by the vanishing of variations of $g$ and $\omega$ with compact support :
\be
S = S[ e,\omega,...] = \int{\Lambda (e,\omega)}
\ee
where $e$ indicate the coframe and $\omega$ indicates the connection, $\Lambda$ is an $n$-form, with $n$ the dimension of the manifold $M$.\\
If we consider the variations with respect to the coframe and the connection we write:
\begin{eqnarray}
\delta_{\omega} S = \int \delta_{\omega} \Lambda \\ \nonumber
\delta_{e} S = \int \delta_{e} \Lambda \\ \nonumber
\end{eqnarray}
and demand:
\begin{eqnarray}
\delta_{\omega} S = 0 \\ \nonumber
\delta_{e} S = 0 \\ \nonumber
\end{eqnarray}
or mod d:
\begin{eqnarray}
\delta_{\omega} \Lambda = 0\\ \nonumber
\delta_{e} \Lambda = 0
\end{eqnarray}
The variational derivative of a tensor $W$ with respect to a generic variable $F$ is indicated by $\underbrace{W}_{F}$ or by $\delta_{F} W$.\\
The variational differentiation of the independent variables like $e$ and $\omega$ will be indicated either by $\dot{e}$, $\dot{\omega}$ or by $\delta e^c$ and $\delta \omega$.\\
The variational differentiation is a type preserving derivation that commutes with contractions and exterior differentiations.\\
\\
In the following after reviewing the so called Divergence conditions, we show how to get the field equations for the generalised Einstein-Hilbert action and then we obtain the Cartan equation for the non-Riemannian part of the connection, and classify some of its solutions.\\
\subsection{Divergence Conditions}
\bigskip
We start with a theory obtained from an action $n$-form which can be written as a sum of $n$-forms $ \sum_{r} \Lambda_{r}$; then in general we find a set of symmetric second degree tensors $\tau_{s}$ which satisfies the condition:\\
\be
\stackrel{o}{\bf \nabla}. \tau_{s} = 0
\ee
In terms of the Levi-Civita connections $\stackrel{o}{\nabla}$.\\
The result relies on the behaviour of the action functional $S$ under the effect of some local diffeomorphisms of the manifold M.\\
We write:
\be
S = \int_{M}{\Lambda}= \sum_{r} \int_{M} \Lambda_{r}
\ee
\bigskip
where $\Lambda$ is constructed in terms of the metric tensor $\bf g$, and other tensor fields $(\phi_{1}, \phi_{2}, \phi_{3}, ....)$. For any local diffeomorphism on $M$ generated by a vector field $V$ we have:
\be
\int_{M} L_{V} \Lambda_{r} = \int_{M} d(i_{V} \Lambda_{r}) = \int_{\partial M} i_{V} \Lambda_{r} = 0
\ee
Since the Lie derivative $L_{V} = i_{V} d + d \, i_{V}$ and $d \Lambda_{r} = 0$ and all fields in $\Lambda_{r}$ are assumed to have compact support.\\
If we consider the variations with respect to the different fields $ (\phi_{1}, \phi_{2},\phi_{3},...)$ which appear in $\Lambda$ we get the equations of motion for $ (\phi_{1}, \phi_{2},\phi_{3},...)$ (mod d), $F_{\phi_{k}} = 0$ where $F_{\phi_{k}}$ are defined by:\\
\be
\underbrace{\Lambda_{r}}_{\phi_{k}} = \dot{\phi_{k}} F_{\phi_{k}}
\ee
In an arbitrary local frame { $X_{a}$ } with dual co-frame { $e^{b}$} such that $ e^{b} (X_{c}) = \delta^{b}_{c}$, we can define the symmetric stress tensor, $\tau_{r} = \tau_{rab} \, e^{a} \otimes e^{b}$.\\
By performing the variation with respect to ${\bf g}$ we get (mod d):
\be
\underbrace{\Lambda_{r}}_{g} = -\dot{g}(X_{a},X_{b}) \tau_{r}^{ab} \star 1
\ee
When the variation is generated by a local diffeomorphism on $M$ by a vector field V then $\dot{\Lambda}_{r} = L_{V} \Lambda_{r}$ and we can write:
\begin{eqnarray}
L_{V} \Lambda_{r} = - (L_{V} {\bf g})(X_{a},X_{b}) \tau^{ab}_{r} \star 1 \\ \nonumber
(L_{V} \phi_{1}) F_{\phi_{1}} + (L_{V} \phi_{2}) F_{\phi_{2}} + ..... 
\end{eqnarray}
then the set $(F_{\phi_{k}})$ is a set of variational equations, if these are satisfied then by using the relation:
\begin{equation}
(L_{V} {\bf g}) (X_{a}, X_{b}) =i_{a} \stackrel{o}{\bf \nabla}_{X_{b}} \mbox{\~{V}} + i_{b} \stackrel{o}{\bf \nabla}_{X_{a}} \mbox{\~{V}}
\end{equation}
where $\mbox{\~{V}} = {\bf g}(V,-)$ and using the symmetry of $\bf g$, it is possible to write:
\begin{equation}
\frac{1}{2} \int_{M} L_{V} {\Lambda}_{r} = - \int_{M} \tau_{r}^{ab} i_{a}\stackrel{o}{\bf \nabla}_{X_{b}} \tilde{V} \star 1
\end{equation}
We can introduce the $(n-1)$ forms $J_{rV}$ by:
\be
J_{rV} = \star \tau_{r}(V,-)
\ee
Now
\be
d(J_{rV}) = d(\tau_{rab}V^{a} \star e^{b})
\ee
The previous expression can be written in the form
\be
(\stackrel{o}{\nabla}.\tau_{r})(V) \star 1 + \tau_{r}^{ab} i_{a} \stackrel{o}{\nabla}_{X_{b}} \tilde{V} \star 1
\ee
where
\be
\stackrel{o}{\nabla}.\tau_{r} \equiv (\nabla_{X_{a}} \tau_{r})(X^{a},-)
\ee
If we consider fields with compact support then:
\begin{eqnarray}
\frac{1}{2} \int_{M} L_{V} \Lambda_{r} = - \int_{M} \tau_{r}^{ab} i_{a} \stackrel{o}{\nabla}_{X_{b}} \tilde{V} \star 1 = \\ \nonumber
- \int_{M} d(J_{rV}) + \int_{M} (\stackrel{o}{\nabla}.\tau_{r})(V) \star 1 = \\ \nonumber
 - \int_{\partial M} J_{rV} + \int_{M}(\stackrel{o}{\nabla}.\tau_{r})(V) \star 1
\end{eqnarray}
We have $J_{V} = 0$ on $\partial M$ and V arbitrary so:
\be
\stackrel{o}{\nabla}.\tau_{r} = 0
\ee
Relation (56) does not represent any ``conservation law''. Conservation laws require something stronger, for example, if a vector $k$ exists (Killing vector) such that $L_{\bf k} {\bf g} =0$ we obtain from (55) that
\be
d J_{r{\bf k}} = 0
\ee
and $J_{r{\bf k}}$ defines a genuine  conserved current.\\
\\
\subsection{Variations of the Generalised\\ Einstein-Hilbert Action}
\bigskip
A general non-Riemannian geometry is specified when we give a metric $\bf g$ and a general connection $\bf \nabla$.\\
In a local coframe ${e^a}$ with dual frame $X_{b}$ such that $e^{a}(X_{b}) = \delta^{a}{}_{b}$, the connection forms satisfy ($\Lambda^{a}{}_{b} \equiv \omega^{a}{}_{b}$):
\be
\omega^{c}{}_{b}(X_{a}) \equiv e^c(\nabla_{X_{a}} X_{b})
\ee
In the following we use orthonormal frames so that:
\be
{\bf g} = \eta_{ab} e^a \otimes e^b
\ee
with $\eta_{ab} = diag(-1,1,1,1,....)$.\\
It is important to observe that the position (59) permits to transfer the functional dependence on the metric $\bf g$ to the coframe $e^{a}$.\\
We consider an action written in the form:
\be
S[{\bf e},{\bf \omega}] = \int \Lambda({\bf e},{\bf \omega})
\ee
for some $n$-form $\Lambda$.\\
The field equations of the theory follow from (mod d):
\begin{eqnarray}
\underbrace{\bf \Lambda}_{\bf e} = 0 \\ \nonumber
\underbrace{\bf \Lambda}_{\bf \omega} = 0 \\ \nonumber
\end{eqnarray}
The general curvature scalar is:
\be
R = i^b i_{a} R^{a}{}_{b}
\ee
The generalised Einstein-Hilbert action density $\Lambda_{EH} = R \star 1$, can be written:
\be
\Lambda_{EH} \equiv R \star 1 = (i^b i_{a} R^{a}{}_{b}) \star 1 = R^{a}{}_{b} \wedge \star (e_{a} \wedge e^b)
\ee
From the definition of curvature two forms it follows that:
\be
\underbrace{R \star 1}_{\omega} = (\underbrace{d \omega^{a}{}_{b} + \omega^{a}{}_{c} \wedge \omega^{c}{}_{b}}_{\omega}) \wedge \star(e_{a} \wedge e^b)
\ee
We can write:
\be
d(\omega^{a}{}_{b} \wedge \star(e_{a} \wedge e^{b})) = d \omega^{a}{}_{b} \wedge \star(e_{a} \wedge e^{b}) - \omega^{a}{}_{b} \wedge d(\star(e_{a} \wedge e^{b}))
\ee
so that:
\begin{eqnarray}
\underbrace{\Lambda_{EH}}_{\omega} = \dot{\omega}^{a}{}_{b} \wedge d \star(e_{a} \wedge e^{b}) + (\underbrace{\omega^{a}{}_{c} \wedge \omega^{c}{}_{b}}_{\omega}) \wedge \star(e_{a} \wedge e^{b}) \\ \nonumber
+ d(\dot{\omega}^{a}{}_{b} \wedge \star(e_{a} \wedge e^{b})) = (\dot{\omega}^{a}{}_{c} \wedge \omega^{c}{}_{b} - \dot{\omega}^{c}{}_{b} \wedge \omega^{a}{}_{c}) \wedge \star (e_{a} \wedge e^{b}) \\ \nonumber
+ {\dot{\omega}}^{a}{}_{b} \wedge d \star (e_{a} \wedge e^{b}) + d(\dot{\omega}^{a}{}_{b} \wedge \star(e_{a} \wedge e^{b})) = \dot{\omega}^{a}{}_{b} \wedge \omega^{b}{}_{c} \wedge \star (e_{a} \wedge e^{c}) \\ \nonumber
- \dot{\omega}^{a}{}_{b} \wedge \omega^{c}{}_{a} \wedge \star (e_{c} \wedge e^{b}) + {\dot{\omega}}^{a}{}_{b} \wedge d \star (e_{a} \wedge e^{b}) + d(\dot{\omega}^{a}{}_{b} \wedge \star(e_{a} \wedge e^{b})) \\ \nonumber
= \dot{\omega}^{a}{}_{b} \wedge D \star(e_{a} \wedge e^{b}) + d(\dot{\omega}^{a}{}_{b} \wedge \star(e_{a} \wedge e^{b}))
\end{eqnarray}
$D$ being the exterior covariant derivative.\\
Since $\dot{\omega}^{a}{}_{b}$ has compact support:
\be
\int \underbrace{{\Lambda}_{EH}}_{\omega} = \int {\dot{\omega}^{a}{}_{b}} \wedge D \star(e_a \wedge e^b)
\ee
\bigskip
The coframe variation gives:
\begin{eqnarray}
\underbrace{{\Lambda}_{EH}}_{e} = \underbrace{R^{a}{}_{b} \wedge \star (e_{a} \wedge e^{b})}_{e} = \\ \nonumber
R^{a}{}_{b} \wedge \underbrace{\star(e_{a} \wedge e^{b})}_{e} = \delta e^{c} \wedge R^{a}{}_{b} \star (e_{a} \wedge e^{b} \wedge e_{c})
\end{eqnarray}
because $R^{a}{}_{b}$ is a coframe independent object.\\
We can write:
\be
\underbrace{\Lambda_{EH}}_{e} = \delta e^{c} \wedge G_{c}
\ee
where the Einstein 3-forms are:
\be
G_{c} = R^{a}{}_{b} \wedge \star (e_{a} \wedge e^b \wedge e_{c})
\ee
We have for any coframe independent $p$ forms $\alpha$ and $\beta$:
\be
\underbrace{\alpha \wedge \star \beta}_{e} = - \dot{e^c} \wedge [i_{c} \beta \wedge \star \alpha - (-1)^p \alpha \wedge i_{c} \star \beta]
\ee
this may be proved as follows:
if we write the generic $p$-form $\beta$ as:
\be
\beta = \beta_{a_{1},a_{2},.....a_{p}} e^{a_{1}a_{2}....a_{p}}
\ee
from the fact $\beta$ is frame independent we get:
\be
\frac{1}{p} \delta_{e}({\beta_{a_{1},a_{2},.....a_{p}}}) e^{a_{1}a_{2}....a_{p}} + \beta_{a_{1},a_{2},.....a_{p}} (\delta_{e} e^{a_{1}} \wedge e^{a_{2}....a_{p}}) = 0
\ee
we have:
\be
\delta_{e}(\star \beta) = \delta_{e}(\beta_{a_{1},a_{2},.....a_{p}} \star e^{a_{1}a_{2}....a_{p}})
\ee
so
\begin{eqnarray}
\underbrace{\alpha \wedge \star \beta}_{e} = \alpha \wedge \delta(\beta_{a_{1},a_{2},.....a_{p}} \star e^{a_{1}a_{2}....a_{p}})\\ \nonumber
= \alpha \wedge \delta_{e}(\beta_{a_{1},a_{2},.....a_{p}}) \star e^{a_{1}a_{2}....a_{p}} + \alpha \wedge \beta_{a_{1},a_{2},.....a_{p}} \delta(\star e^{a_{1}a_{2}....a_{p}})\\ \nonumber
= \delta_{e}(\beta_{a_{1},a_{2},.....a_{p}})e^{a_{1}a_{2}....a_{p}} \wedge \star \alpha + (-1)^p \delta e^c \wedge \alpha \wedge \beta_{a_{1},a_{2},.....a_{p}} \star(e^{a_{1}a_{2}....a_{p}}{}_{c}) \\ \nonumber
= - p \beta_{a_{1},a_{2},.....a_{p}} \delta e^{a_{1}} \wedge e^{a_{2}....a_{p}} \wedge \star \alpha +  (-1)^p \delta e^c \wedge \alpha \wedge i_{c} \star \beta \\ \nonumber
= - \delta e^c \wedge [i_{c} \beta \wedge \star \alpha - (-1)^p \alpha \wedge i_{c} \star \beta]
\end{eqnarray}
where use has been made of the property $\alpha \wedge \star \beta = \beta \wedge \star \alpha$ in the third line and of (73) in the 4th line.\\
From expression (71) we can get a relation which relates the scalar curvature to the stress forms. To obtain this relation consider the fact that if $\alpha$ and $\beta$ are frame dependent then we have to add to (71) the coframe variation of $\alpha$ and $\beta$, let us define $\Delta \tau_{c}$ by:
\be
\delta e^c \wedge \Delta \tau_{c} = \underbrace{\alpha}_{e} \wedge \star \beta + \alpha \wedge \star \underbrace{\beta}_{e}
\ee
Consider an action of the form:
\be
\Lambda = k R \star 1 +  \sum_{k} b_{k}(\alpha_{k} \wedge \star \beta_{k})
\ee
where $\alpha_{k}$ and $\beta_{k}$ are generic $p_{k}$-forms and $b_{k}$ are constants.\\
The coframe variation of (77) gives the equations:
\be
k R^{a}{}_{b} \wedge i_{c} i^b i_{a} \star 1 - \sum_{k} b_{k}[i_{c} \beta_{k} \wedge \star \alpha_{k} - (-1)^{p_{k}} \alpha_{k} \wedge i_{c} \star \beta_{k}] + \sum_{k} \Delta \tau_{c}[k] = 0
\ee
where $\Delta \tau_{c}[k]$ is the extra term in the stress forms coming from the generic term $ b_{k}(\alpha_{k} \wedge \star \beta_{k})$.\\
If we consider the wedge product of (78) with $e^c$ we get:
\be
(-1)^{n+1}(n-2) R \star 1 + (-1)^n \sum_{k} [(2p_{k}-n) b_{k}(\alpha_{k} \wedge \star \beta_{k}) + \sum_{k} (\Delta \tau_{c}[k]) \wedge e^c = 0
\ee
\bigskip
Consider now a situation in which we have an action density of the form:
\be
\Lambda_{EH} + F({\bf e}, {\bf \omega})
\ee
The connection variation gives the equation:
\be
D \star (e^a \wedge e_b) = F^a{}_b
\ee
with $F^a{}_b$ $n-1$ forms defined by:
\be
\underbrace{F}_{\omega} =\dot{\omega}^b{}_a \wedge F^{a}{}_{b}
\ee
Equation (81) is called the Cartan equation.\\
The previous equation implies:
\be
F^a{}_{a} = 0
\ee
We can define the set of 0-forms $f^{ca}{}_{b}$ by
\be
f^{ca}{}_{b} = i^{c} \star F^{a}{}_{b} 
\ee
From (83) we get $f^{ca}{}_{a} = 0 $.\\
It is possible to obtain the general solution of the Cartan equation by decomposing $Q_{ab}$ into the trace part and the trace-free part [29]:
\be
Q_{ab} = {\hat{Q}}_{ab} + \frac{1}{n}g_{ab} Q
\ee
in such a way ${\hat Q}^{a}{}_{a} = 0$, and decomposing the torsion in the same way as:
\be
T^a = {\hat{T}}^{a} + \frac{1}{n-1} e^a \wedge T
\ee
Where $T \equiv i_{a} T^a$ and $i_a {\hat{T}}^{a} = 0$\\
Equation (81) can be decomposed as:
\be
i_{b}\hat{Q}_{a}{}^{c} - \delta^{c}{}_{b} i_{d} \hat{Q}_{a}{}^{d} + (\delta^{c}{}_{b} \delta^{d}{}_{a} - \delta^{c}{}_{a}\delta^{d}{}_{b})(\frac{n-2}{2n}i_{d}Q - i_{d}i_{h}T^{h}) - i_{b}i_{a}T^{c} + f^{c}{}_{ab} = 0
\ee
that is,
\begin{eqnarray}
i_{a} \hat{Q}_{bc} - i_{a}i_{b} T_{c} & = & - \frac{1}{2n} g_{bc} i_{a} Q + \frac{1}{2n} g_{ac} i_{b} Q - f_{cba} \\ \nonumber
& - & \frac{1}{n(n-2)} g_{ac} f^{d}{}_{db} + \frac{n-1}{n(n-2)} g_{bc} f^{d}{}_{da} \\ \nonumber
& + & \frac{n-1}{n(n-2)} g_{ac} f^{d}{}_{bd} - \frac{1}{n(n-2)} g_{bc} f^{d}{}_{ad}
\end{eqnarray} 
We obtain using the symmetry of ${\hat{Q}}_{ab}$ and the antisymmetry of $i_{a}i_{b}T_c$ that:
\be
i_a i_b {\hat{T}}_{c} = \frac{1}{n-1}(g_{bc} f^d{}_{ad} - g_{ac} f^{d}{}_{bd}) -\frac{1}{2}(f_{bac} + f_{bca} + f_{cab} - f_{cba} - f_{abc} - f_{acb})
\ee
\be
i_{a} {\hat{Q}}_{bc} = \frac{1}{n} g_{bc}(f^{d}{}_{da} + f^{d}{}_{ad})- \frac{1}{2}(f_{bac}+f_{bca} + f_{cab} + f_{cba} - f_{abc} - f_{acb})
\ee 
and
\be
T - \frac{n-1}{2n}Q = \frac{1}{n(n-2)}(f^{c}{}_{ac} + (1-n)f^{c}{}_{ca})e^a
\ee
Observe the following expression for the trace free part of the torsion:
\be
\hat{T_{c}} = \frac{1}{n-1}(e_{c} \wedge e^{a})f^{d}{}_{ad} - \frac{1}{2}(e^b \wedge e^a)(f_{bac} + f_{bca} + f_{cab})
\ee
The equations (89-92) provide the general solution of the Cartan equation (81), the study of the properties of the general solution is very important for what we will discuss in the next chapter so we will devote an entire paragraph to the presentation of some particular but important cases of the above mentioned equations.\\
Let us observe that in general the connection variation of a generic action gives a Cartan equation which is a differential equation to be solved for the non-Riemannian part of the connection. In the present case, however the fact that we are considering an action density like (80) allows us to solve algebraically eq. (81) for $\lambda^{a}{}_{b}$.\\
\\
\subsection{Particular Solutions of the Cartan Equation}
\bigskip
In this section we present some particular solutions of the Cartan equation (81). The different cases are classified depending on the form assumed by the $n-1$ forms $F^{a}{}_{b}$.\\
We will prove that if $F^{a}{}_{b}$ assumes the particular form of subsection 3.3.6 or 3.3.8 the traceless part of the torsion ${\hat{T}}^{a}$ turns out to be zero. This result will be used in the following chapters.\\
\subsubsection{$F^{a}{}_{b}=0$}
The first case we are going to consider is when $F^{a}{}_{b}=0$.\\
In that case using the equations of the previous paragraph we get:
\be
i_{a} {\hat{Q}}_{bc} = 0
\ee
that is $ {\hat{Q}}_{bc} = 0$ and
\be
i_{a} i_{b} {\hat{T}}_{c} = 0
\ee
which means $ {\hat{T}}_{c} = 0$. We can write:
\be
T^a = \frac{1}{n-1}(e^a \wedge T)
\ee
\be
T = \frac{n-1}{2n}Q
\ee
The non-metricity 1-forms and the torsion 2-forms result to be:
\be
Q_{ab} = \frac{1}{n} g_{ab} Q
\ee
\be
T^a = \frac{1}{2n}(e^a \wedge Q)
\ee
The non-Riemannian part of the connection assumes the following very simple form:
\be
{\lambda}_{ab} = - \frac{1}{2n} Q g_{ab}
\ee
So the traceless part of the non-Riemannian part of the connection defined by:
\be
{\hat{\lambda}}^{a}{}_{b} = {\lambda}^{a}{}_{b} - \frac{1}{n}{\lambda}^{c}{}_{c}\delta^{a}{}_{b} 
\ee
is zero. \\
We will see in the next chapter that this condition guarantees that the non-Riemannian part of the Einstein 3-forms vanish.\\
\\
\subsubsection{$f_{cab} = -f_{cba}$}
\bigskip
In this case by using the formulas of the previous paragraph we get:
\be
{\hat{Q}}_{ab} = 0
\ee
\be
T - \frac{n-1}{2n}Q = \frac{1}{n-2}f^{c}{}_{ac}e^{a}
\ee
The traceless part of the torsion 2-forms is calculated to be
\\
\be
{\hat{T_c}} = \frac{1}{n-1}(e_{c} \wedge e^{a})f^{d}{}_{ad} + \frac{1}{2}(e^{b} \wedge e^{a})f_{cba}
\ee
So the solution can be written:
\begin{equation}
Q_{ab} = \frac{1}{n} g_{ab} Q 
\end{equation}
\be
T_c = \frac{1}{n-1}(e_{c} \wedge e^{a})f^{d}{}_{ad} + \frac{1}{2}(e^b \wedge e^{a})f_{cba} + \frac{1}{n-1}e_{c} \wedge T 
\end{equation}
\\
\subsubsection{$F^{a}{}_{b} = \sum_{k} (e^{a} \wedge i_{b} \star A_{k})$}
\bigskip
In this section we consider the case in which the forms $F^{a}{}_{b}$ can be written as:
\be
F^{a}{}_{b} = \sum_{k} (e^{a} \wedge i_{b} \star A_{k})
\ee
With $A_{k}$ set  of 1-forms.\\
We get easily that:
\be
f_{cab} = [\sum_{k}( i_{c}(A_{k}) g_{ab} - i_{a}(A_{k}) g_{cb})]
\ee
so that:
\be
f_{cab} = - f_{acb}
\ee
We see that:
\be
f^{c}{}_{ac} = [(1-n)i_{a} \sum_{k} A_{k}]
\ee
and
\be
f^{c}{}_{ca} = 0
\ee
The condition $f^{ca}{}_{a} = 0$ gives $(n-1)i^{c} \sum_{k} A_{k} = 0$ that is $\sum_{k} A_{k} = 0$.\\
Then we get:
\be
T = (n-1)(\frac{Q}{2n} - \frac{\sum_{k}(A_{k})}{n(n-2)})
\ee
The calculation of $Q_{bc}$ gives:
\be 
Q_{bc}  =[\sum_{k} (-e_{b} i_{c} A_{k} - e_{c} i_{b} A_{k} + \frac{n+1}{n} g_{bc} A_{k})] + \frac{1}{n} g_{bc} Q
\ee
Indeed from the general expression (90) using the antisymmetry of $f_{cab}$ we have:
\be
Q_{bc} = \frac{1}{n} g_{bc} \, f^{d}{}_{ad} e^a + f_{abc} e^a + f_{acb} e^a + \frac{1}{n} Q g_{bc}
\ee
using the expression (107) we have:
\begin{eqnarray}
f^{d}{}_{ad} e^a = (1-n)\sum_{k} A_{k}\\ \nonumber
f_{acb} e^a = [\sum_{k}(A_{k} g_{cb} - i_{c}(A_{k}) e_b)]\\ \nonumber
f_{abc} e^a = [\sum_{k}(A_{k} g_{cb} - i_{b}(A_{k}) e_c)]\\ \nonumber
\end{eqnarray}
using the previous relations we get the result for $Q_{bc}$. Analogously using the relation (92) we get:
\begin{eqnarray}
\hat{T}_{c} & = & [\sum_{k}(\frac{1}{n-1}(e_c \wedge e^a)(1-n) i_{a}(A_{k})\\ \nonumber 
& + & \frac{1}{2}(e^b \wedge e^a)[i_{a} A_{k} g_{bc} - i_{b}A_{k} g_{ac}+ i_{c} A_{k} g_{bc} \\ \nonumber 
& - & i_{b} A_{k} g_{ac} + i_{a}A_{k} g_{bc} - i_{c}A_{k} g_{ab}])\\ \nonumber
& = & \sum_{k}(- (e_{c} \wedge A_{k}) + e^b \wedge (e^a \wedge i_{a} A_{k})g_{bc} + (e^a \wedge e^b) i_{b} (A_{k}) g_{ac})]= \sum_{k}( e_{c} \wedge A_{k}) \\ \nonumber
\end{eqnarray}
So the traceless part of the torsion comes to be:
\\
\be
\hat{T_{c}} = [e_{c} \wedge \sum_{k}(A_{k})]
\ee
\\
\subsubsection{$F^{a}{}_{b} = {\delta}^{a}{}_{b} \star A$}
\bigskip
With $A$ a generic 1-form in this case:
\be
f_{cab} =  g_{ab} i_{c} A
\ee
The condition $f^{ca}{}_{a} = 0$ gives $n \, i^c A = 0$.\\
We observe that:
\be
T - \frac{n-1}{2n} Q = - \frac{A}{n}
\ee
Using the same method of the previous case we get:
\be
\hat{T_{c}} = \frac{n}{n-1}(e_{c} \wedge A)
\ee
so that:
\be
T_{c} = \frac{n+1}{n}(e_{c} \wedge A) + \frac{1}{2n} (e_{c} \wedge Q)
\ee
The non-metricity can be calculated using the expression (90) which using the symmetry properties of $f_{cab}$ becomes:
\be
i_{a}{\hat Q}_{bc} = \frac{2}{n}g_{bc}f^{d}{}_{da} - f_{bac} - f_{cab} + f_{acb}
\ee
so that:
\be
{\hat{Q}}_{bc} = \frac{2}{n}g_{bc}f^{d}{}_{da} e^a - f_{bac}e^a - f_{cab}e^a + f_{acb}e^a
\ee
By plugging in the expression for $f_{abc}$ we get:
\begin{eqnarray}
{\hat{Q}_{bc}} & = & [\frac{2}{n}g_{bc}i_{a}(A)e^a - g_{ac} (i_{b}A) e^a\\ \nonumber
 & - & g_{ab} (i_{c}A) e^a + g_{cb} (i_{a}A) e^a\\ \nonumber
 & = & \frac{2}{n} g_{bc} A - i_{b} A e_c - i_{c} A e_b + g_{bc} A\\ \nonumber
 & = &-e_{c}i_{b}A - e_{b}i_{c}A + \frac{2+n}{n}g_{bc}A\\ \nonumber
\end{eqnarray}
So we get:
\be
Q_{bc} = [-e_{b} i_{c}(A) - e_{c} i_{b}(A) + \frac{n+2}{n} g_{bc} A] + \frac{1}{n} Q g_{bc}
\ee
The last two cases will be used in chapter 4. As we will see in the next chapter they will help to understand the relation between certain models of non-Riemannian gravity and Einstein theory.
\\
\subsubsection{$F^{a}{}_{b} = e^a \wedge \star A_{b},_{}i^{b} A_{b} = 0$}
\bigskip
where $A_{b}$ is a two form. In this case we require $A_{b}$ to be traceless because if not traceless it can be written as:
\be
A_{b} = \hat{A_{b}} + \frac{1}{n-1}(e_{b} \wedge A)
\ee
with $ A = i^{a} A_{a}$. The contribution from the second term gives a term of the type treated in 3.3.3, so we limit to consider the traceless case.\\
We find easily that:
\be
f_{cab} = i_{a}i_{c} A_{b}
\ee
so that:
\begin{eqnarray}
f_{cab} = - f_{acb}\\ \nonumber
f^{ca}{}_{a} = i^{c}i^{a}(A_{a}) = 0 \\ \nonumber
f^{cc}{}_{a} = 0 \\ \nonumber
f^{ca}{}_{c} = i^{a}i^{c}(A_{c}) = 0
\end{eqnarray} 
So in this case we get:
\be
T - \frac{n-1}{2n} Q = 0
\ee
In this case the relation between $Q$ and $T$ is the same as relation (96).\\
After some calculation we find:
\be
{\hat{Q}}_{ab} = - [i_{a}(A_{b}) + i_{b}(A_{a})]
\ee
and
\be
\hat{T_{c}} = - [e^{a} \wedge i_{c}(A_{a}) + A_{c}]
\ee
which means
\be
e^{c} \wedge \hat{T_{c}} = e^a \wedge A_{a}
\ee
If we consider the case in which $A_{b}$ can be written as:
\be
A_{b} = i_{b} B
\ee
with $B$ a 3-form then $\hat{Q}_{ab} = 0.$\\
Another important case which we will meet in the next chapter, is when $A_{b}$ coincides itself with $\hat{T}_{b}$ times a constant $\lambda$: $A_{b} = \lambda \hat{T}_{b}$, then from relation (131) we get that if:
\be
\lambda = 1
\ee
then $\hat{T}_{c}$ is arbitrary while if $\lambda \neq 1$ we need:
\be
e^c \wedge \hat{T}_{c} = 0
\ee
\subsubsection{(case 3.3 + 3.4) $F^a{}_{b} = {\delta}^{a}{}_{b} \star A + \sum_{k} (e^a \wedge i_{b} \star A_{k})$}
\bigskip
We get:
\be
f_{cab} = \sum_{k} ((i_{c}A_{k}) g_{ab} - (i_{a} A_{k})g_{cb}) + g_{ab}(i_{c}A)
\ee
The condition $f^{ca}{}_{a} = 0$ gives:
\be
(n-1)\sum_{k} A_{k} + n A = 0
\ee
We get then:
\be
T = \frac{n-1}{2n} Q - [\frac{n-1}{n(n-2)} \sum_{k} A_{k} + \frac{1}{n} A]
\ee
The non-metricity is:
\be
Q_{bc} =  [- e_{b} i_{c} (\sum_{k}A_{k} + A) -  e_{c} i_{b} (\sum_{k}A_{k} + A) + \frac{n+1}{n} g_{bc} \sum_{k} A_{k} + \frac{n+2}{n} g_{bc} A] + \frac{1}{n} Q g_{bc}
\ee
using relation (136) we get:
\be
Q_{bc} = [-e_{b} i_{c}(\sum_{k} A_{k} + A) - e_{c} i_{b} (\sum_{k} A_{k} + A) + \frac{2}{n} (\sum_{k} A_{k} + A)g_{bc}] + \frac{1}{n}g_{bc} Q
\ee
By defining $A_{1} = - (\sum_{k} A_{k} + A)$ we can write:
\be
Q_{bc} = e_{b} i_{c} A_{1} + e_{c} i_{b} A_{1} - \frac{2}{n} A_{1}g_{bc}  + \frac{1}{n}g_{bc} Q
\ee
The traceless part of the torsion is by using (116,119) and (136):
\be
\hat{T}_{c} =  [e_{c} \wedge \sum_{k} A_{k} + \frac{n}{n-1} e_{c} \wedge A] = 0
\ee
so that:
\be
T^c = \frac{1}{n-1} (e^c \wedge T)
\ee
\\
\subsubsection{ $F^{a}{}_{b} = \sum_{k} (e_{b} \wedge i^{a} \star A_{k})$}
\bigskip
In this case we get:
\be
f_{cab} = [\sum_{k}(i_{c}(A_{k})g_{ab})- i_{b}(A_{k})g_{ac} ]
\ee
so that:
\be
f_{cab} = - f_{bac}
\ee
we see that $f^{c}{}_{ac} = 0$ and:
\be
f^{c}{}_{ca} = [(1-n)i_{a} \sum_{k} A_{k}]
\ee
the condition $f^{ca}{}_{a} = 0 $ gives $(n-1)i^{c} \sum_{k} A_{k} = 0$
we then get:
\be
T - \frac{n-1}{2n}Q =  \sum_{k}[\frac{(n-1)^2}{n(n-2)}A_{k}]
\ee
the calculation of $Q_{bc}$ give the same expression we found for the case 3.3.3:
\be
Q_{bc} = [\sum_{k}(-e_b i_{c} A_{k} - e_{c}i_{b} A_{k} + \frac{n+1}{n}g_{bc}A_{k})] + \frac{1}{n} g_{bc}Q
\ee
as far as the traceless part of the torsion is concerned we get:
\be
\hat{T}_{c} = \frac{1}{2}(e^b \wedge e^a) f_{bca} = - (e^b \wedge e^a) g_{ca} \sum_{k} i_{b} A_{k} = e^c \wedge \sum_{k} A_{k}
\ee
that is the same expression of the case 3.3.3\\
\\
\subsubsection{(case 3.7 + 3.4) $F^{a}{}_{b} = \delta^{a}{}_{b} \star A + \sum_{k}(e_{b} \wedge i^a \star A_{k})$}
\bigskip
We have:
\be
f_{cab} = [\sum_{k}(i_{c}(A_{k})g_{ab} - i_{b}(A_{k})g_{ac}) + g_{ab} i_{c} A
\ee
all the results are equivalent to the case 3.3.6, apart from the relation between $T$ and $Q$ which is modified into:
\be
T = \frac{n-1}{2n}Q + [\frac{(n-1)^2}{n(n-2)} A_{k} - \frac{A}{n}]
\ee
The expression for $Q_{bc}$ and $T_{c}$ are the same and the traceless part of the torsion 2-forms comes out to be zero. 
\newpage
\begin{center}
\bf CHAPTER 4
\end{center}
\section{NON-RIEMANNIAN GRAVITATION AND THE ISOMORPHISM WITH GENERAL RELATIVITY}
\bigskip

In this chapter we investigate the remarkable relation which exists between certain models of non-Riemannian gravity and Einstein's theory, this relation was first discovered by Tucker, Wang [22,25-26] two years ago, but they limited their analysis to the relation with Einstein-Proca models, and the proof they presented for the general $n$ dimensional case was computer based.\\
Here we reconsider the problem and in several steps we give a proof by hand of a theorem which was first proved with the help of the computer language Maple concerning the simplification of the field equations. We manage to do so dividing the proof in different simple but significant steps, by doing so it becomes clear how the proof can be easily extended to other theories like scalar tensor gravities, then we are able to formulate a more general result which can be applied to more general cases, as we will see, this will permit us in the next chapter to obtain a class of new black-hole solutions with non-Riemannian connections and scalar fields, so far no such solutions have been found in the literature.\\
We start the analysis by studying the action which was first considered in Ref. [25] where they considered only $n=4$; we start by considering the case in which the Weyl 1-form $Q$ satisfies the equation $d \star dQ = 0$. \\
Here we present a step by step proof valid for any $n$, then we generalise the proof to the case the equation for $Q$ contains a mass term and the action from which we get the equations of motion is more general.\\
As we will see we may consider quite general actions which give for the Weyl 1-form $Q$ the Proca type equation $\alpha \, d \star d Q + \beta \, \star Q = 0$. When this happens we say that the model is 'Proca-type'.\\
The existence of the remarkable simplification will bring us to the definition of $\bf Isomorphism$ between certain non-Riemannian models of gravity and Levi-Civita theories, this being the fact that after the simplification the field equations for the non-Riemannian gravity are equivalent to the Levi-Civita one, with the non-Riemannian fields $(Q,T,...)$ being replaced by certain fields which appear in the Levi-Civita case. This relation can be used to obtain solutions for the non-Riemannian field equations using the solutions of the Levi-Civita theory.\\
Using the formalism of chapter 2 and 3 we present the study of the conformal generalised Einstein equations that is the case where we have a scalar field coupled with the Einstein-Hilbert term $f(\psi) \, R \star 1$ where $\psi$ is a dilaton field.\\
In section 4.5 we consider the case in which the term $f(\psi) \, R \star 1$ does not appear in the action. Properties of Isomorphism may be shown to hold; these will be used in the next chapter to get a new class of black hole solutions with non-Riemannian connections.
\subsection{Relation with Einstein-Proca theory: massless Weyl field case}

\bigskip
We start the analysis by considering the relation with the Einstein-Proca models in the case in which the mass of the Weyl field associated with the non-metricity 1-forms is zero, this case though simpler will be used in the following to get all the other results.\\
To begin with, consider the action [25]:
\bigskip
\be
 \int \Lambda[{\bf e}, {\bf \omega}] = \int  k R \star 1 + \frac{\alpha}{2} dQ \wedge \star dQ + \frac{\beta}{2} Q \wedge \star Q + \frac{\gamma}{2} T \wedge \star T
\ee
\bigskip
Where $k, \alpha , \beta , \gamma$ are real constants, $R, Q, T$ are the curvature scalar the trace of the non-metricity 1-forms, and the contracted forms $T = i_{c}T^c$, defined in chapter 2.\\
To get the field equations we need to consider the variation with respect to the coframe ${e^a}$ and the connection ${\omega}^a{}_{b}$.\\
The first thing to note is that using relation (17) of Chapter 2 we get that:
\be
\underbrace{Q}_{\bf e} = 0
\ee
and
\be
\underbrace{dQ}_{\bf e} = d(\underbrace{Q}_{\bf e}) = 0
\ee
since $Q$ is a frame independent object.\\
On the other hand the coframe variation of $T^a$ and $T$ in general are different from zero.\\
Using the formula (71) we get:\\
\be
\underbrace{\frac{1}{2}(dQ \wedge \star dQ)}_{\bf e} = \delta e^a \wedge \frac{1}{2}(dQ \wedge i_{a} \star dQ - i_{a}dQ \wedge \star dQ)
\ee
\be
\underbrace{\frac{1}{2}(Q \wedge \star Q)}_{\bf e} = - \delta e^a \wedge \frac{1}{2}(Q \wedge i_{a} \star Q + i_{a}Q \wedge \star Q)
\ee
The variation of the Einstein-Hilbert term has already been considered in the previous chapter and it gives:
\be
\delta e^c \wedge R^{a}{}_{b} \wedge \star(e_a \wedge e^b \wedge e_c)
\ee
The calculation of the variation of the term containing the torsion requires a bit more formalism, for instance the coframe variation of $T$ is:
\be
\delta_e T = \delta_e (i_{a} T^a) = \delta_e (i_{a}) T^a + i_{a} (\delta_e T^a)
\ee
Let us consider the first term in the previous relation. We need to obtain some properties for the operator $\delta_e (i_{a})$ which in the following will be indicated as:
\be
i_{\delta_{a}}
\ee
The first thing to observe is that using the definition of the coframe $i_{a}(e^b) = \delta^{b}{}_{a}$ and $\delta_{e} (\delta^{b}{}_{a}) = 0$ we get:
\be
i_{\delta_{a}} e^b + i_{a} (\delta e^b) = 0
\ee
or:
\be
i_{\delta_{a}} e^b = - i_{a} (\delta e^b)
\ee
using this property it follows that $i_{\delta_{a}}$ satisfies the same properties of an ordinary interior derivation, for example:
\be
i_{\delta_{a}}(e^b \wedge e^c) =  i_{\delta_{a}}(e^b) \wedge e^c - e^b \wedge i_{\delta_{a}}(e^c)
\ee
by means of the previous property we can prove a simple but useful relation.\\
Let us consider a generic $p$-form $B$, we can write $p B = e^b \wedge i_{b} B$, so that
\begin{eqnarray}
p i_{\delta_{a}} B = i_{\delta_{a}}(e^b) \wedge i_{b} B - e^b  \wedge i_{\delta_{a}} i_{b} B = \\ \nonumber
i_{\delta_{a}}(e^b) \wedge i_{b} B  + e^b \wedge i_{b} i_{\delta_{a}} B = (p-1) i_{\delta_{a}} B + i_{\delta_{a}}(e^b)  \wedge i_{b} B
\end{eqnarray}
from which we get:
\be
i_{\delta_{a}} B = i_{\delta_{a}}(e^c) \wedge i_{c} B
\ee
this relation proves useful in calculating variations since while in general we do not know the effect of $i_{\delta_{a}}$ on a generic $p$-form $B$, by using relation (160) we know the effect on the coframe vectors $e^c$.\\
We can use the above relation for calculating one of the contributions in the variation of the torsion term:
\begin{eqnarray}
\underbrace{\frac{1}{2}(T \wedge \star T)}_{\bf e} & = & \underbrace{T}_{\bf e} \wedge \star T + \frac{1}{2}(T \wedge \underbrace{\star}_{\bf e} T) \\ \nonumber
& = &(i_{\delta_{a}} T^a + i_{a} (\delta T^a)) \wedge \star T - \delta e^a \wedge \frac{1}{2}(T \wedge i_{a} \star T + i_{a}T \wedge \star T) \\ \nonumber
& = & i_{\delta_{a}}(e^b) \wedge i_{b}(T^a) \wedge \star T - \delta T^a \wedge i_{a} \star T - \delta e^a \wedge \frac{1}{2}(T \wedge i_{a} \star T \\ \nonumber
& + & i_{a} T \wedge \star T) = - i_{a} \delta e^b \wedge i_{b}(T^a) \wedge \star T + \delta e^b \wedge \lambda^{a}{}_{b} \wedge i_{a} \star T \\ \nonumber
& - & \delta e^a \wedge \frac{1}{2}(T \wedge i_{a} \star T + i_{a} T \wedge \star T) = i_{a} \delta e^b \wedge T^a \wedge i_{b} \star T \\ \nonumber
& + & \delta e^b \wedge \lambda^a{}_{b} \wedge i_{a} \star T - \delta e^a \wedge \frac{1}{2}(T \wedge i_{a} \star T + i_{a} T \wedge \star T) \\ \nonumber
& = & \delta e^b  \wedge i_{a}(T^a \wedge i_{b} \star T) + \delta e^b \wedge \lambda^a{}_{b} \wedge i_{a} \star T \\ \nonumber 
& - & \delta e^a \wedge \frac{1}{2}(T \wedge i_{a} \star T + i_{a} T \wedge \star T)
\end{eqnarray}
so the stress forms associated with the torsion term is:
\be
\tau_{a}[\gamma] = \gamma (i_{b}(T^b \wedge i_{a} \star T) + \lambda^b{}_{a} \wedge i_{b} \star T - \frac{1}{2}(T \wedge i_{a} \star T + i_{a} T \wedge \star T))
\ee
let us move now to the connection variations. As we know from the previous chapter the connection variation of the Einstein-Hilbert term is (mod d):
\be
\underbrace{R \star 1}_{\bf \omega} = \dot{\omega}^{a}{}_{b} \wedge D \star(e_{a} \wedge e^b)
\ee
to calculate the variation of the other terms we observe that:
\be 
\underbrace{Q}_{\bf \omega} = g^{ab} \underbrace{Q_{ab}}_{\bf \omega} = g^{ab} \underbrace{Dg_{ab}}_{\bf \omega} = -2 \dot{\omega}^a{}_{b} \delta^{b}{}_{a}
\ee
\be
\underbrace{T^a}_{\bf \omega} = \dot{\omega}^{a}{}_{b} \wedge e^b
\ee
then we get
\be
\frac{1}{2} \underbrace{Q \wedge \star Q}_{\bf \omega} = -2 \dot{\omega}^{a}{}_{b} \delta^{b}_{a} \wedge \star Q
\ee
and mod d
\be
\frac{1}{2} \underbrace{dQ \wedge \star dQ}_{\bf \omega} = -2 \dot{\omega}_{a}^{b} \delta^{a}{}_{b} \wedge d \star dQ
\ee
The term containing the torsion gives:
\be
\frac{1}{2} \underbrace{T \wedge \star T}_{\bf \omega} =  i_{a} \underbrace{T^a}_{\bf \omega} \wedge \star T = 
i_{a}(\dot{\omega}^{a}{}_{c} \wedge e^c) \wedge \star T = - \dot{\omega}^{a}{}_{b} \wedge e^b \wedge i_{a} \star T
\ee
thus the variation with respect to the connection gives mod d the field equation
\be
k D \star (e_{a} \wedge e^b) = 2\delta^{b}{}_{a}(\alpha d \star d Q + \beta \star Q) + \gamma e^b \wedge i_{a} \star T
\ee
by taking the trace of this equation it is possible to replace it by the other two equivalent equations:
\be
\alpha d \star d Q + \beta \star Q  = \frac{\gamma (1-n)}{2n} \star T
\ee
\be
k D \star (e_{a} \wedge e^b) = \delta^{b}{}_{a} \frac{1-n}{n} \gamma \star T + \gamma e^b \wedge i_{a} \star T
\ee
where $n$ is the dimension of the manifold.
The generalised Einstein equations which come from the coframe variation are:
\be
k R^{a}{}_{b} \wedge \star (e_{a} \wedge e^b \wedge e_{c}) + \tau_{c}[\alpha] + \tau_{c}[\beta] + \tau_{c}[\gamma] = 0
\ee
where
\be
\tau_{c}[\alpha] = \frac{\alpha}{2}(dQ \wedge i_{c} \star dQ - i_{c} dQ \wedge \star dQ)
\ee
\be
\tau_{c}[\beta] = -\frac{\beta}{2}(Q \wedge i_{c} \star Q + i_{c} Q \wedge \star Q)
\ee
\be
\tau_{c}[\gamma] = \gamma [i_{k}(T^k \wedge \star(T \wedge e_{c})) + \lambda^{k}{}_{c} \wedge i_{k} \star T - \frac{1}{2} (T \wedge i_{c} \star T + i_{c}T \wedge \star T)]
\ee
Let us solve equation (172) for the non-Riemannian part of the connection.\\
We can rewrite equations (174) as:
\be
D \star (e^b \wedge e_{a}) = \delta^{b}{}_{a} \frac{n-1}{n} \frac{\gamma}{k} \star T - \frac{\gamma}{k} e^b \wedge i_{a} \star T
\ee
apart from the exchange of $a$ and $b$ this equation corresponds to the case 3.3.6 of the Cartan equation discussed in the previous chapter, with the correspondence:
\be
A = \frac{n-1}{n} \frac{\gamma}{k}T
\ee
\be
\sum_{k} A_{k} = -\frac{\gamma}{k} T
\ee
so we can use  relation (137) to get:
\be
T - \frac{n-1}{2n} Q = [\frac{1-n}{n^2} \frac{\gamma}{k} T + \frac{n-1}{n(n-2)} \frac{\gamma}{k}T] =
[\frac{\gamma}{k} T \frac{2(n-1)}{n^2(n-2)}]
\ee
or
\be
T = [\frac{n-1}{2n} \frac{Q}{1- \frac{2 \gamma (n-1)}{k n^2(n-2)}}]
\ee
\bigskip
Observe the interesting fact that $T$ is directly proportional to the trace of the non-metricity  1-form $Q$.\\
Indeed by using the relations of the previous chapter we have now proved the following:\\
\\
{\bf Proposition} : \emph{For theories in which $F^{a}{}_{b}$ takes the form $F^{a}{}_{b} = \delta^{a}{}_{b} \star A + \sum_{k}(e^{a} \wedge i_{b} \star A_{k})$, with $A_{k} = C_{1k}T + C_{2k}Q$ and $ A = C_{3}T + C_{4}Q$ with constants $C_{1k},C_{2k},C_{3},C_{4}$ it follows that, $T = C_{5}Q$ for some constant $C_{5}$}.\\
\\
Such theories will be called 'Proca-type'.\\
This name comes from the fact that if we plug (183) into (173) we obtain the Proca equation $\alpha d \star dQ + \beta_{0} \star Q = 0$ where:
\begin{eqnarray*}
\beta_{0} = \beta - \frac{1}{4}\frac{\gamma k \, (n-1)^2 (n-2)}{(k \, n^2 (n-2) - 2 \gamma (n-1))}
\end{eqnarray*}
\bigskip
If we require the Weyl field to be massless i.e. we demand $d \star dQ = 0$ then from equation (173):
\be
Q = \frac{\gamma (1-n)}{\beta 2 n} T
\ee
using relation (182) we get:
\be
T[1 + \frac{(n-1)^2 \gamma}{4 \beta n^2} - \frac{2 \gamma (n-1)}{k n^2 (n-2)}] = 0
\ee
which is equivalent to (since we are supposing $T \neq 0$):
\be
4 n^2 (n-2) \beta k + (n-1)^2 (n-2) \gamma k + 8 (1-n) \gamma \beta = 0
\ee
In the particular case of $n=4$:
\be
64 \beta k + 9 \gamma k - 12 \beta \gamma = 0
\ee
so the coupling constants would have to be constrained if we want to have a massless Weyl field.\\
For the 'Proca-type' theories using the results of case 3.3.6 of previous the chapter we conclude that the traceless part of the torsion is zero so that we can write:
\be
T^a = \frac{1}{n-1}(e^a \wedge T)
\ee
To calculate the non-Riemannian part of the connection we use the relation (23) in which the non-metricity 1-forms are given by formula (140).\\  
After a simple calculation we find that:
\be
\lambda^{a}{}_{b} = \frac{1}{2n}[ e^a \wedge i_{b}(A_{3} + 2 A_{1} - A_{2}) + e_{b} \wedge i^{a}(A_{2} - (2+2n)A_{1} - A_{3}) + \delta^{a}{}_{b}(2A_{1} - A_{2})]
\ee
where $A_{3} = \frac{2n}{n-1}T$, $A_{2} = Q$, and $A_{1} = -(A + \sum_{k}A_{k})$.\\
The traceless part of $\lambda^{a}{}_{b}$ is:
\be
\hat{\lambda}^{a}{}_{b} = \frac{1}{2n}[ e^a \wedge i_{b}(A_{3} + 2 A_{1} - A_{2}) + e_{b} \wedge i^{a}(A_{2} - (2+2n)A_{1} - A_{3}) + \delta^{a}{}_{b} 2 A_{1}]
\ee
from relation (182) we get:
\be
\frac{1}{2n}(A_{3} - A_{2}) = \frac{\gamma}{k}T \frac{2}{n^2 (n-2)}
\ee
since $A_{1} = -[\sum_{k} A_{k} + A]$ we get:
\be
A_{1} = \frac{1}{n} \frac{\gamma}{k} T
\ee
and
\be
A_{3} - A_{2} = \frac{4}{n-2} A_{1}
\ee
so we may rewrite the non-Riemannian part of the connection in the form:
\be
\lambda^{a}{}_{b} = [\frac{1}{n-2} i_{b}(A_{1}) \wedge e^a + \frac{1-n}{n-2} e_{b} \wedge i^{a}(A_{1}) + \frac{1}{2n} \delta^{a}{}_{b}(2A_{1} - A_{2})]
\ee
We can observe that in order to get the non-Riemannian part of the connection 1-forms we have used only the traceless part of the Cartan equation (174). By using (174) we can obtain $\lambda^{a}{}_{b}$ as a function of the torsion $T$. If we introduce in the action extra terms which modifies (173) but not (174) then the algebraic dependence of $\lambda^{a}{}_{b}$ as a function of $T$ will be the same. So we realize that the torsion has somehow a special role in determining the non-Riemannian part of the connection; the origin of this resides in the fact the variations considered in (169-170) contain the diagonal operator $\delta^{a}{}_{b}$ which (171) does not contain.\\
This observation will be used in section 4.5 to generalise the cancellation property of the field equations to some dilaton gravity models.\\
In the following we want to use the expression for $\lambda^{a}{}_{b}$ to calculate all the terms which appear in the generalised Einstein equations. As we will see, a remarkable cancellation occurs in the latter equations which permits us to simplify the study of the properties of non-Riemannian actions.\\
The expression for $\lambda^{a}{}_{b}$ can be used to calculate the term $\lambda^{b}{}_{a} \wedge i_{b} \star T$:
\begin{eqnarray}
\lambda^{b}{}_{a} \wedge i_{b} \star T = \\ \nonumber
 [\frac{1}{n-2} i_{a}(A_{1}) \wedge e^b \wedge i_{b} \star T 
+ \frac{1-n}{n-2}[i^{b}(A_{1}) \wedge e_{a} \wedge i_{b} \star T] + \\ \nonumber
\frac{1}{2n}[2A_{1} - A_{2}] \wedge i_{a} \star T = \frac{n-1}{n-2}[i_{a}(A_{1}) \wedge \star T ] \\ \nonumber
+ \frac{1-n}{n-2}[i^{b}(A_{1}) \wedge e_{a} \wedge i_{a} \star T] + \frac{1}{2n}(2A_{1} - A_{2}) \wedge i_{a} \star T] 
\end{eqnarray}
Then using the relation:
\be
i^{k} A \wedge e_{a} \wedge i_{k} \star T = - i_{k}[i^{k} A \wedge \star i_{a} T] + i_{a} A \wedge \star T
\ee
we can rewrite the expression in the form:
\be
\lambda^{b}{}_{a} \wedge i_{b} \star T = [\frac{n-1}{n-2} i_{b}[i^{b}(A_{1}) \wedge \star i_{a}T] + \frac{1}{2n}[2A_{1} - A_{2}] \wedge i_{a} \star T]
\ee
By using the definition of $A_{3}$ we can write the term $i_{k}(T^k \wedge \star(T \wedge e_{a}))$ as:
\be
i_{k}(T^k \wedge \star(T \wedge e_{a})) = \frac{1}{2n} A_{3} \wedge i_{a} \star T
\ee
Adding this term to the relation (197) and using (193) gives:
\begin{eqnarray}
i_{b}(T^b \wedge i_{a} \star T) + \lambda^{b}{}_{a} \wedge i_{b} \star T = \\ \nonumber
[\frac{n-1}{n-2} i_{b}(i^b(A_{1}) \wedge \star i_{a} T) + \frac{1}{2n}[2 A_{1} - A_{2} + A_{3}] \wedge i_{a} \star T = \\ \nonumber
\frac{n-1}{n-2} i^b [i_{b}(A_{1}) \wedge \star i_{a} T] + \frac{1}{n-2} A_{1} \wedge i_{a} \star T]
\end{eqnarray}
The $ \star$ variation of the torsion term gives:
\be
- \frac{1}{2}(T \wedge i_{a} \star T + i_{a} T \wedge \star T) = \frac{1-n}{4n} [A_{3} \wedge i_{a} \star T + i_{a} A_{3} \wedge \star T]
\ee
The stress forms which come from the term $Q \wedge \star Q$ can be written:
\be
- \frac{1}{2}(Q \wedge i_{a} \star Q + i_{a}Q \wedge \star Q) = - \frac{1}{2}(A_{2} \wedge i_{a} \star T + i_{a}A_{2} \wedge \star T) \lambda
\ee
where $\lambda$ is defined by:
\be
Q = \lambda T
\ee
If the Weyl field mass is zero then from equation (184) we get:
\be
\lambda = \frac{\gamma}{\beta} \frac{1-n}{2n}
\ee
so that
\be
- \frac{1}{2}(Q \wedge i_{a} \star Q + i_{a}Q \wedge \star Q) = - \frac{\gamma}{2 \beta} \frac{1-n}{2n}(A_{2} \wedge i_{a} \star T + i_{a}A_{2} \wedge \star T)
\ee
Let us move now to the calculation of the non-Riemannian contribution to the Einstein tensor. Using the formula (34) of chapter 2 we can write the Einstein-Hilbert term as:
\begin{eqnarray}
R \star 1 = \stackrel{o}{R} \star 1 - \lambda^{a}{}_{c} \wedge \lambda^{c}{}_{b} \wedge \star(e^b \wedge e_{a}) - d(\lambda^{a}{}_{b} \wedge \star(e^b \wedge e_{a}))\\ \nonumber
= \stackrel{o}{R} \star 1 - \hat{\lambda}^{a}{}_{c} \wedge \hat{\lambda}^{c}{}_{b} \wedge \star(e^b \wedge e_{a}) - d(\hat{\lambda}^{a}{}_{b} \wedge \star(e^b \wedge e_{a}))
\end{eqnarray}
where ${\hat{\lambda}}^{a}{}_{b}$ indicates the traceless part of the non-Riemannian part of the connection.\\
The non-Riemannian contribution to the Levi-Civita Einstein-Hilbert term which we indicate by $\Delta \stackrel{o}{R} \star 1$ is mod d:
\be
\Delta \stackrel{o}{R} \star 1 = \hat{\lambda}^{a}{}_{c} \wedge \hat{\lambda}^{c}{}_{b} \wedge \star(e_{a} \wedge e^b)
\ee
its coframe variation gives:
\be
\tau_{d}(\Delta \stackrel{o}{R} \star 1) = \hat{\lambda}^{a}{}_{c} \wedge \hat{\lambda}^{c}{}_{b} \wedge \star(e_{a} \wedge e^b \wedge e_{d})
\ee
by using relation (194) we get:
\be
\tau_{a}(\Delta \stackrel{o}{R} \star 1) = \frac{1}{2-n}[A_{1} \wedge i_{a} \star A_{1} + i_{a} A_{1} \wedge \star A_{1}] - n \, i^b \, [i_{b}(A_{1}) \wedge \star i_{a}(A_{1})]
\ee
it is convenient in the following to express the torsion as a function of $A_{1}$, using relation (192) we can write:
\be
\tau_{a}[\beta] = \frac{n-1}{4}[A_{2} \wedge i_{a} \star A_{1} + i_{a} A_{2} \wedge \star A_{1}]
\ee
and
\begin{eqnarray}
\tau_{a}[\gamma] = \\ \nonumber
k \frac{1-n}{4}[A_{3} \wedge i_{a} \star A_{1} + i_{a} A_{3} \wedge \star A_{1}] \\ \nonumber
+ k \frac{n(n-1)}{n-2} i^b[i_{b}(A_{1}) \wedge \star i_{a}(A_{1})] + \\ \nonumber
k \frac{n}{n-2}(A_{1} \wedge  i_{a} \star A_{1})
\end{eqnarray}
we get then:
\begin{eqnarray}
\tau_{a}[\beta] + \tau_{a}[\gamma] + \tau_{a}(\Delta \stackrel{o}{R} \star 1) = \\ \nonumber
k \frac{n-1}{4}[A_{2} \wedge i_{a} \star A_{1} + i_{a} A_{2} \wedge \star A_{1}] + \\ \nonumber 
k \frac{1-n}{4}[A_{3} \wedge i_{a} \star A_{1} + i_{a} A_{3} \wedge \star A_{1}] \\ \nonumber
+ k \frac{n(n-1)}{n-2} i^b[i_{b}(A_{1}) \wedge \star i_{a}(A_{1})] + \\ \nonumber
k \frac{n}{n-2}(A_{1} \wedge  i_{a} \star A_{1}) + k \frac{1}{2-n}[A_{1} \wedge i_{a} \star A_{1} + \\ \nonumber
i_{a} A_{1} \wedge \star A_{1}] - k \, n \, i^b[i_{b}(A_{1}) \wedge \star i_{a}(A_{1})]
\end{eqnarray}
\newpage
using (193) we find:
\begin{eqnarray}
\tau_{a}[\beta] + \tau_{a}[\gamma] + \tau_{a}(\Delta \stackrel{o}{R} \star 1) = \\ \nonumber
k [\frac{1}{2-n}[A_{1} \wedge i_{a} \star A_{1} + i_{a}A_{1} \wedge \star A_{1}] + \frac{n}{n-2}i^b[i_{b}(A_{1}) \wedge \star i_{a}(A_{1})] \\ \nonumber
+ \frac{n}{n-2}(A_{1} \wedge i_{a} \star A_{1}) + \frac{n-1}{2-n} [A_{1} \wedge i_{a} \star A_{1} + i_{a}A_{1} \wedge \star A_{1}]] = \\ \nonumber
k [\frac{n}{2-n}[A_{1} \wedge i_{a} \star A_{1} + i_{a}A_{1} \wedge \star A_{1}] + \frac{n}{n-2}[A_{1} \wedge i_{a} \star A_{1}] + \\ \nonumber
\frac{n}{n-2}i^b[i_{b} A_{1} \wedge \star i_{a} A_{1}]] = k [\frac{n}{2-n}[i_{a} A_{1} \wedge \star A_{1} - i^b[i_{b} A_{1} \wedge \star i_{a} A_{1}]]] = 0
\end{eqnarray}
where use has been made of the relation:
\begin{eqnarray*}
i_{a} A \wedge \star A = i^{b}[i_{b} A \wedge \star i_{a} A]
\end{eqnarray*}
for a generic 1-form $A$.\\
The generalised Einstein equations then simplifies into:
\be
k \stackrel{o}{G}_{c} + \tau_{c}[\alpha]  = 0
\ee
where ${\stackrel{o}{G}}_{c} = \stackrel{o}{R^{a}{}_{b}} \wedge \star (e_{a} \wedge e^{b} \wedge e_{c})$, we conclude that the non-Riemannian contribution to the Einstein tensor exactly cancels all the stress terms apart from $\tau_{c}[\alpha]$ and the equations (175) simplify into (213).\\ 
This is the cited simplification originally discovered by Tucker, Wang in the case $n=4$ [25].\\
This implies that if we consider a generic solution of the Einstein-Maxwell system, then provided relation (186) is satisfied we get a solution of the equations (172,175) of the action (151) with the non-metricity given by (140),
and
\be
A_{2} = A_{EM}
\ee
\be
A_{1} = \frac{\gamma}{nk}T = \frac{2 \beta}{(1-n)k} A_{EM}
\ee
\be
A_{3} = A_{2} + \frac{4}{n-2}A_{1}
\ee
\bigskip
\subsection{Massive Weyl field}
\bigskip
Suppose now that the Weyl field has mass parameter $\beta_{0}$ so that:
\be
\alpha d \star d Q + \beta_{0} \star Q = 0
\ee
from equation (173) we have:
\be
(\beta - \beta_{0}) \star Q = \frac{\gamma (1-n)}{2 n} \star T
\ee
so
\be
\lambda = \frac{\gamma (1-n)}{2n} \frac{1}{\beta - \beta_{0}} =  \frac{\gamma (1-n)}{2n \beta}(1+ \frac{\beta_{0}}{\beta - \beta_{0}})
\ee
then (209) is replaced by:
\be
\tau_{a}[\beta] = k \frac{n-1}{4}[1 +\frac{\beta_{0}}{\beta - \beta_{0}}][A_{2} \wedge i_{a} \star A_{1} + i_{a} A_{2} \wedge \star A_{1}]
\ee
then an extra term appears which is not cancelled out:
\be
\Delta \tau_{a}[\beta] = k \frac{n-1}{4}[\frac{\beta_{0}}{\beta - \beta_{0}}][A_{2} \wedge i_{a} \star A_{1} + i_{a} A_{2} \wedge \star A_{1}]
\ee
which using (192) and (218) can be rewritten as:
\be
-\beta_{0} \frac{1}{2}[Q \wedge i_{a} \star Q + i_{a} Q \wedge \star Q]
\ee
then the Einstein equations reduce to:
\be
k \stackrel{o}{G}_{c} + \tau_{c}[\alpha] + \tau_{c}[\beta_{0}] = 0
\ee
That is the Einstein equations for the Proca field with mass parameter $\beta_{0}$.\\ 
Relation (186) becomes:
\be
4 n^2 (n-2) (\beta - \beta_{0}) k + (n-1)^2 (n-2) \gamma k + 8 (1-n) \gamma (\beta - \beta_{0}) = 0
\ee
The forms $A_{1},A_{2},A_{3}$ become:
\be
A_{2} = A_{EP}
\ee
\be
A_{1} = \frac{\gamma}{nk}T = \frac{2 (\beta - \beta_{0})}{(1-n)k} A_{EP}
\ee
and:
\be
A_{3} = A_{2} + \frac{4}{n-2}A_{1}
\ee
\\
We can summarise paragraphs 4.1 and 4.2 by stating the following:\\
\\
{\bf Theorem} : \emph{The solutions of the field equations which come from the action (151) also satisfy the Einstein-Proca system (equations (217) and (223) in a pseudo-riemannian geometry)}. 
\subsection{Extended Actions}
\bigskip
In this section we extend what found in the previous two sections to show that the simplification property holds even for more general actions.\\
Consider now the theory obtained from the action:
\be
\int \Lambda[{\bf e}, {\bf \omega}] = \int [k R \star 1 + \frac{\alpha}{2} dQ \wedge \star dQ + \frac{\beta}{2} Q \wedge \star Q + \frac{\gamma}{2} T \wedge \star T + \frac{\epsilon}{2}(T^c \wedge \star T_{c})]
\ee
where $T^c$ is the torsion two form. \\
By considering the coframe variation we get:
\be
\tau_{a}[\epsilon] = \epsilon( - \lambda^{b}{}_{a} \wedge \star T_{b} + \frac{1}{2}(T_{b} \wedge i_{a} \star T^{b} - i_{a} T_{b} \wedge \star T^{b}))
\ee
while the Cartan equation yields:
\be
\alpha d \star dQ + \beta \star Q = \frac{\gamma (1-n)}{2n} \star T + \frac{\epsilon}{2n} e^c \wedge \star T_{c}
\ee
\be
k D \star (e_{a} \wedge e^b) = \delta^{b}{}_{a} \frac{(1-n)}{n} \gamma \star T + \delta^{b}{}_{a}\frac{\epsilon}{n} e^{c} \wedge \star T_{c} + \gamma e^b \wedge i_{a} \star T - \epsilon e^b \wedge \star T_{a}
\ee
by using relation (20) we can write these equations in the form:
\be
\alpha d \star dQ + \beta \star Q = \frac{1}{2n}[\gamma (1-n) - \epsilon] \star T
\ee
\be
k D \star (e_{a} \wedge e^b) = \frac{\delta^{b}{}_{a}}{n}[\gamma (1-n) - \epsilon] \star T + \gamma e^b \wedge i_{a} \star T - \epsilon e^b \wedge \star T_{a}
\ee
The last term in the second equation can be written:
\be
\epsilon (e^b \wedge \star T_{a}) = \epsilon (e^b \wedge \star[{\hat{T}}_{a} + \frac{1}{n-1}(e_{a} \wedge T)]) = \epsilon (e^b \wedge \star {\hat{T}}_{a}) + \frac{\epsilon}{1-n}(e^b \wedge i_{a} \star T) 
\ee
where $\hat{T}_{a}$ denotes the traceless part of the torsion satisfying $i^{a}\hat{T}_{a} = 0$.\\
We can write equation (233) as:
\be
k D \star (e_{a} \wedge e^b) = \frac{\delta^{b}{}_{a}}{n}[\gamma (1-n) - \epsilon] \star T + \frac{1}{1-n}[\gamma (1-n) - \epsilon] e^b \wedge i_{a} \star T - \epsilon e^b \wedge \star {\hat{T}}_{a}
\ee
equations (232) and (235) are equivalent apart from the last term to equations (173,174) with $\gamma \rightarrow \gamma - \frac{\epsilon}{1-n}$.\\ 
If we define $\gamma'$ as:
\be
\gamma '  = \gamma - \frac{\epsilon}{1-n}
\ee
We can write:
\be
\alpha d \star d Q + \beta \star Q = \frac{\gamma ' (1-n)}{2n} \star T
\ee
\be
k D \star(e_{a} \wedge e^b) = \delta^{b}{}_{a}\frac{\gamma ' (1-n)}{n} \star T + \gamma ' e^b \wedge i_{a} \star T - \epsilon(e^b \wedge \star {\hat{T}}_{a})
\ee
\bigskip
We can apply now the formulas (129,130) of the third chapter to get the extra contribution to the traceless part of the non-metricity 1-forms ${\hat{Q}}_{ab}$, we call it $ \Delta{\hat{Q}}_{ab}$: 
\be
\Delta {\hat{Q}}_{ab} = - \epsilon [i_{a} \hat{T}_{b} + i_{b} \hat{T}_{a}]
\ee
Using relation (128) we see that the linear relation between $T$ and $Q$ remains the same moreover:
\be
e^c \wedge \hat{T}_{c} = \frac{\epsilon}{k}(e^c \wedge \hat{T}_{c}) 
\ee
From the last one it follows that if we put:
\be
\epsilon = k
\ee
then $\hat{T}_{c}$ is \emph{arbitrary}.\\
If instead $\epsilon \neq k$ then:
\be
e^c \wedge \hat{T}_{c} = 0
\ee
Consider firstly the case $\hat{T}_{a} = 0$. The constants $\epsilon$ and k are not constrained, but as we will see there is still a constraint between $\gamma '$,  $\beta$ and $k$.\\
If $\hat{T}_{a} = 0$ the Cartan equation (237,238) coincides with (173-174) with $\gamma'$ defined in (236), then we can use the expression for the non-Riemannian part of the connection (194) in which $A_{1}$ is given by:
\be
A_{1} = \frac{\gamma' T}{nk}
\ee
relation (224) becomes:
\be
4 n^2 (n-2) (\beta - \beta_{0}) k + (n-1)^2 (n-2) \gamma' k + 8 (1-n) \gamma' (\beta - \beta_{0}) = 0
\ee
The forms $A_{2}, A_{3}$ are given by the same relations (225,227).\\
Let us move now to the stress forms, since the traceless part of the torsion two form vanishes then:
\be
\star T_{a} = \frac{1}{1-n}i_{a} \star T
\ee
we get then:
\be
T_{c} \wedge i_{a} \star T^{c} = \frac{2-n}{(n-1)^2}(T \wedge i_{a} \star T)
\ee
and
\be
i_{a}T_{c} \wedge \star T^c = -\frac{1}{(n-1)^2}(T \wedge i_{a} \star T) + \frac{1}{n-1}(i_{a}T \wedge \star T)
\ee
so that:
\be
\frac{1}{2}(T_{c} \wedge i_{a} \star T^{c} - i_{a} T_{c} \wedge \star T_{c}) = 
\frac{3-n}{2(n-1)^2}(T \wedge i_{c} \star T) - \frac{1}{2(n-1)} (i_{a}T \wedge \star T)
\ee
it is easily checked that:
\be
i_{k}(T^k \wedge \star(T \wedge e_{a})) - \frac{1}{2}(T \wedge i_{a} \star T + i_{a}T \wedge \star T) = \frac{3-n}{2(n-1)}(T \wedge i_{a} \star T) - \frac{1}{2}(i_{a} T \wedge \star T)
\ee
we see that using relation (245) we have:
\be
-\epsilon(\lambda^{b}{}_{a} \wedge \star T_{b}) = - \frac{\epsilon}{1-n} \lambda^{b}{}_{a} \wedge i_{b} \star T
\ee
so the result is that:
\be
\tau_{a}[\epsilon] = -\frac{\epsilon}{1-n}\tau_{a}[\gamma]
\ee
from this follows that:
\be
\tau_{a}[\gamma] + \tau_{a}[\epsilon] = \tau_{a}[\gamma - \frac{\epsilon}{1-n}] = \tau_{a}[\gamma']
\ee
The conclusion is that if we choose the traceless part of the torsion 2-form equal to zero, the action (228) is equivalent to (151) with the replacement:
\be
\gamma \rightarrow \gamma'
\ee
so the cancellation proved for the action (151) is valid for the action (228) too.\\
It is possible to prove using the same technique used in the previous paragraphs that the cancellation still occurs when ${\hat{T}^c} \neq 0$.\\
\\
Consider now a theory obtained from an action density:
\be
\Lambda[{\bf e}, {\bf \omega}] = k R \star 1 + \frac{\alpha}{2} dQ \wedge \star dQ + \frac{\beta}{2} Q \wedge \star Q + \frac{\gamma}{2} T \wedge \star T + \frac{\nu}{2}(Q \wedge \star T)
\ee
the connection variation of the term $Q \wedge \star T$ gives:
\be
\underbrace{(Q \wedge \star T)}_{\omega} = -2 \dot{\omega}^{a}{}_{b} \delta^{a}{}_{b} \wedge \star T -  \dot{\omega}^{a}{}_{b} \wedge e^b \wedge i_{a} \star Q
\ee
the terms which appear in the former expression are of the type considered in 3.3.6 of the previous chapter, we can use relation (137) to conclude that in this case too the torsion 1-form $T$ is proportional to the trace of the non-metricity 1 - form $Q$:
\be
T = \lambda Q
\ee
where $\lambda$ can be calculated using the equations of chapter 3.\\ 
\bigskip
The Cartan equation yields:
\be
\bigskip
\alpha d \star dQ + \beta \star Q + \frac{1}{2} \nu \star T = (\gamma + \frac{1}{2}\nu \mu)\frac{1-n}{2n} \star T
\ee
\be
k D \star (e_{a} \wedge e^b) = \delta^b{}_{a} (\gamma + \frac{1}{2}\nu \mu) \frac{1-n}{n} \star T + (\gamma + \frac{1}{2}\nu \mu) e^b \wedge i_{a} \star T
\ee
\bigskip
\bigskip
where $\mu = \frac{1}{\lambda}$.
Eq. (257) can be rewritten:
\be
\alpha d \star dQ + \beta \star Q + \frac{1}{2}\nu \lambda \star Q = (\gamma + \frac{1}{2}\nu \mu)\frac{1-n}{2n} \star T
\ee
then we have:
\be
\frac{1}{2} \underbrace{\nu(Q \wedge \star T)}_{e} = - \frac{1}{2}\nu[i_{c} T \wedge \star Q + Q \wedge i_{c} \star T] + \frac{1}{2}\nu \underbrace{T}_{e} \wedge \star Q 
\ee
which can be written as:
\be
- \frac{1}{2} \nu \lambda \frac{1}{2}[i_{c}Q \wedge \star Q + Q \wedge i_{c} \star Q] + \frac{1}{2} \nu \mu \underbrace{T}_{e} \wedge \star T - \frac{1}{2} \nu \mu \frac{1}{2}[i_{c} T \wedge \star T + T \wedge i_{c} \star T]
\ee
which is:
\be
\frac{1}{2} \mu \nu \frac{1}{2} \underbrace{(T \wedge \star T)}_{e} - \frac{1}{2}\nu \lambda \frac{1}{2}[i_{c} Q \wedge \star Q + Q \wedge i_{c} \star Q]
\ee
the conclusion is that the inclusion of the term $\frac{\nu}{2} (Q \wedge \star T)$ is equivalent to:
\be
\gamma \rightarrow \gamma + \frac{1}{2}\nu \mu
\ee
\be
\beta \rightarrow \beta + \frac{1}{2}\nu \lambda
\ee
the value of $\lambda$ can be calculated using the formulas of the previous chapter.\\ 
The contribution to the Einstein 3-form can be calculated using the formula (207) with:
\bigskip
\be
A_{1} = (\gamma + \frac{\nu \mu}{2})\frac{T}{nk}
\ee
\bigskip
The conclusion is that apart from a redefinition of the coupling constants, the reduction of the field equations to the Einstein-Proca will occur in this case too.\\
This property can be generalised to more general actions, in particular actions which contain:
\be
{\cal{Q}} = e_{a} i_{b} Q^{ab}
\ee
Let us consider first the term:
\be
{\cal{Q}} \wedge \star T
\ee
The connection variation is:
\begin{eqnarray}
\underbrace{{\cal{Q}} \wedge \star T}_{\omega} = 
\underbrace{{\cal{Q}}}_{\omega} \wedge \star T + \underbrace{T}_{\omega} \wedge \star {\cal{Q}} \\ \nonumber
= \underbrace{e_{a} i_{b} Q^{ab}}_{\omega}  \wedge \star T - \dot{\omega}^a_{b} \wedge e^b \wedge i_{a} \star {\cal{Q}} = \underbrace{Q^{ab}}_{\omega} \wedge i_{b}[e_{a} \wedge \star T] \\ \nonumber
- \dot{\omega}^a_{b} \wedge e^b \wedge i_{a} \star {\cal{Q}} = -(\underbrace{\lambda^{ab} + \lambda^{ba}}_{\omega}) \, g_{ab} \wedge \star T + \underbrace{\lambda^{ab} + \lambda^{ba}}_{\omega} \wedge e_{a} \wedge i_{b} \star T \\ \nonumber
- \dot{\omega}^a{}_{b} \wedge e^b \wedge i_{a} \star {\cal{Q}} = -2 \dot{\omega}^a{}_{b} \delta^a_{b} \wedge \star T - \dot{\omega}^{a}{}_{b} \wedge e^b \wedge i_{a} \star {\cal{Q}} \\ \nonumber
+ \dot{\omega}^a{}_{b} \wedge e_{a} \wedge i^{b} \star T + \dot{\omega}^a{}_{b} \wedge  e^b \wedge i_{a} \star T
\end{eqnarray}
Analogously we get for a ${\cal{Q}} \wedge \star Q$ the variation:
\be
\underbrace{{\cal{Q}} \wedge \star Q}_{\omega} = -2 {\dot{\omega}}^a{}_{b} \delta^a{}_{b} \wedge \star (Q + {\cal{Q}}) + {\dot{\omega}}^a{}_{b} \wedge e_{a} \wedge i^{b} \star Q + {\dot{\omega}}^a{}_{b} \wedge e^{b} \wedge i_{a} \star Q
\ee
and 
\be
\underbrace{{\cal{Q}} \wedge \star {\cal{Q}}}_{\omega} = - 4 {\dot{\omega}}^a{}_{b} \delta^a{}_{b} \wedge \star {\cal{Q}} + 
2 {\dot{\omega}}^a{}_{b} \wedge e_{a} \wedge i^b \star {\cal{Q}} + 2 {\dot{\omega}}^a{}_{b} \wedge e^{b} \wedge i_{a} \star {\cal{Q}}
\ee
We can see that all the terms which appear in the previous formulas are of the type considered in chapter 3 in our study of the Cartan equation.\\ 
By using the formulas in there we can say that $T$ and ${\cal{Q}}$  are directly proportional to the trace of non metricity 1-forms $Q$, it is very easy to prove that the field $Q$ will satisfy a Proca-type equation, then using the same techniques of the previous paragraphs we can verify that the cancellation still occurs in the field equations.\\
We will see in the next sections that it is possible to extend this property of simplification to actions which contain more general terms, but before proceeding let us introduce the concept of {\bf Isomorphism} between a non-Riemannian theory and Levi-Civita theories. We consider the case in which the Levi-Civita action is dependent on the metric $\bf g$ and on a 1-form $A$. The generalisation to more general cases is straightforward.\\
\\
{\bf Definition}: \emph{We say there is an isomorphism between a non-Riemannian theory obtained from an action $\Lambda_{NR}({\bf g}, \omega)$, and a Levi-Civita theory obtained from an action $\Lambda_{LC}({\bf g}, A)$, if the field equations obtained from the action $\Lambda_{NR}$ coincide with the field equations obtained from the action $\Lambda_{LC}$, provided we replace the non-Riemannian fields with certain functions obtained from the field $A$}.\\
\\
{\bf Example}: The theory obtained from the action (151) is isomorphic to the theory obtained from the Levi-Civita action:
\be
\Lambda_{LC} = k \stackrel{o}{R} \star 1 + \frac{\alpha}{2}(dA \wedge \star dA) + \frac{\beta_{0}}{2}(A \wedge \star A)
\ee
for a generic 1- form $A$, because the field equations we obtain from the latter action considering the variation with respect to the metric and the 1-form $A$, are the same as equations (217,223) provided we replace $A$ by $Q$. This implies that we can use solutions of the field equations to the theory (271) to get solutions for the field equations of (151), (217,223).\\
In the next section we present a case in which the isomorphism property does not occur; namely the case in which we have a Dilaton field conformally coupled to the full non-Riemannian Einstein-Hilbert term.\\
\bigskip
\subsection{Conformal Generalised Einstein Equations}
\bigskip
In this section we study the effect on the generalised Einstein equations of the term $\psi^2 \, R \star 1$.\\
Consider the action:
\be
\int \Lambda[e,\omega,\psi] = \int k (1 + \alpha \psi^2) R \star 1 + \frac{\beta}{2}(d \psi \wedge \star d \psi)
\ee
where $\psi$ is a scalar field (0-form), and $R$ is the full non-Riemannian scalar curvature, $\alpha, \beta$ are constants.\\
The variation of the action with respect to $\psi$ gives:
\bigskip
\be
\beta d \star d \psi - 2 k  \, \alpha \, \psi \, (R \star 1) = 0
\ee
\bigskip
The calculation of the connection variation proceeds in a similar way to eq. (67) but now instead of the term $d \omega^{a}{}_{b} \wedge \star (e^{b} \wedge e_{a})$  we have the term:
\be
(1+ \alpha \psi^2) \, (d \omega^{a}{}_{b} \wedge \star (e^{b} \wedge e_{a}))
\ee
we have:
\begin{eqnarray}
[d \omega^{a}{}_{b} \wedge \star (e_{a} \wedge e^{b})](1 + \alpha \psi^2) = \\ \nonumber
d[(1+ \alpha\psi^2)[\omega^{a}{}_{b} \wedge \star (e_{a} \wedge e^{b})]] + \omega^{a}{}_{b} \wedge d[\star(e_{a} \wedge e^{b})(1 + \alpha \psi^2)]
\end{eqnarray}
so (mod d):
\begin{eqnarray}
d \omega^{a}{}_{b} \wedge \star (e_{a} \wedge e^{b})(1+ \alpha \psi^2) = \\ \nonumber
\omega^{a}{}_{b} \wedge d[\star(e_{a} \wedge e^{b})(1+ \alpha \psi^2)] = \\ \nonumber
\omega^{a}{}_{b} \wedge (1+ \alpha \psi^2) \, d(\star(e_{a} \wedge e^{b})) + 2 \, \alpha \, \psi \, (\omega^{a}{}_{b} \wedge d\psi \wedge \star (e_{a} \wedge e^{b}))
\end{eqnarray}
Then analogously to equation (67) we get the Cartan equation:
\be
(1+ \alpha \psi^2) \, D \star (e_{a} \wedge e^{b}) + 2 \alpha \psi[d \psi \wedge \star (e_{a} \wedge e^{b})] = 0
\ee
or
\be
D \star (e_{a} \wedge e^b) = A(\psi)[d \psi \wedge \star(e_{a} \wedge e^b)]
\ee
where:
\be
A(\psi) = -\frac{2 \alpha \psi}{1+ \alpha \psi^2}
\ee
To calculate the coframe variation of the term $(R \star 1)(1 + \alpha \psi^2)$ using the decomposition (205), observe that:
\bigskip
\begin{eqnarray}
d[(1 + \alpha \psi^2)(\hat{\lambda}^a{}_{b} \wedge \star(e^b \wedge e_{a}))] = \\ \nonumber
(1 + \alpha \psi^2) d(\hat{\lambda}^a{}_{b} \wedge \star(e^b \wedge e_{a})) - 2 \alpha \psi \wedge \dot{\lambda}^a{}_{b} \wedge d \psi \wedge \star (e^b \wedge e_{a})
\end{eqnarray}
\bigskip
so (mod d):
\bigskip
\be
d(\hat{\lambda}^a{}_{b} \wedge \star (e^b \wedge e_{a}))(1 + \alpha \psi^2) = 2 \, \alpha \, \psi [\hat{\lambda}^a{}_{b} \wedge d \psi \wedge \star(e^b \wedge e_{a})]
\ee
\bigskip
the coframe variation then gives the Einstein equations:
\begin{eqnarray}
k (1 + \alpha \psi^2) \stackrel{o}{R^{a}{}_{b}} \wedge \star (e^b \wedge e_{a} \wedge e_{c}) + \\ \nonumber
- 2 \alpha \psi [\hat{\lambda}^a{}_{b} \wedge d \psi \wedge \star (e^b \wedge e_{a} \wedge e_{c})] - \frac{\beta}{2} (d \psi \wedge i_{c} \star d \psi + i_{c} d \psi \wedge \star d \psi) \\ \nonumber
+ k(1+ \alpha \psi^2)({\hat{\lambda}}^{a}{}_{c} \wedge {\hat{\lambda}}^{c}{}_{b}) \wedge \star (e_{a} \wedge e^{b} \wedge e_{c}) = 0
\end{eqnarray}
\bigskip
to continue the analysis we need to solve the Cartan equation for $\lambda^{a}{}_{b}$:
\bigskip
\be
D \star (e_{a} \wedge e^b) = A(\psi)[d \psi \wedge \star(e_{a} \wedge e^b) ]
\ee
we find:
\bigskip
\be
f_{cab} = A(\psi) i_{c}(\star(d \psi \wedge \star(e_{a} \wedge e_{b})))
\ee
\bigskip
we see that $f_{cab} = - f_{cba}$ so from the results of chapter 3 we can say that the traceless part of the non-metricity 1- forms vanish:
\be
\hat{Q}^{ab} = 0
\ee
We can write $f_{cab}$ as:
\be
f_{cab} = A(\psi) \psi_{d}[\delta^d_{b} g_{ca} - \delta^d_{a} g_{cb}]
\ee
where the 0-forms $\psi_{d}$ are defined by $d \psi = \psi_{d}e^d$.\\
Contracting (286) we find:
\be
f^{c}{}_{ac} = A(\psi) \psi_{d} (1-n) \delta^{d}{}_{a}
\ee
the two latter expressions may be used to calculate the traceless part of the torsion which turns out to vanish:
\be
\hat{T}_{c} = 0
\ee
so that $T^a = \frac{1}{n-1} e^a \wedge T$.\\
Then since $f^{c}{}_{ac} = - f^{c}{}_{ca}$ applying relation (91) we get:
\be
T = \frac{n-1}{2n}Q + \frac{1-n}{n-2}A(\psi) d \psi
\ee
The solution for the non-metricity and torsion can then be written as:
\bigskip
\begin{eqnarray}
Q_{ab} = \frac{1}{n}g_{ab} Q \\ \nonumber
T^a = \frac{1}{2n} e^a \wedge Q - \frac{1}{n-2}(e^a \wedge d \psi) A(\psi)
\end{eqnarray}
\bigskip
using the relation (23) of chapter 2 we get the following expression for the non-Riemannian part of the connection 1-forms:
\be
\lambda_{ab} =- \frac{1}{2n} g_{ab} Q + \frac{1}{n-2}A(\psi)(i_{a}(d \psi) e_{b} - i_{b}(d \psi) e_{a})
\ee
and the traceless part:
\be
\hat{\lambda}_{ab} = \frac{1}{n-2}A(\psi)(i_{a}(d \psi) e_{b} - i_{b}(d \psi) e_{a})
\ee
Let us consider the terms which appear in the generalised Einstein's equations.\\ Consider first the term:
\bigskip
\be
(2 \alpha \psi) [\hat{\lambda}^{a}{}_{b} \wedge d \psi \wedge \star(e^b \wedge e_{a} \wedge e_{c})]
\ee
Using the expression for the non-Riemannian part of the connection 1-forms we get:
\be
4 \alpha \psi A(\psi)[i^{a}(d \psi) \wedge d \psi \wedge i_{a} i_{c} \star 1]
\ee
which can be written as:
\bigskip
\be
4 \alpha \psi A(\psi)[i_{c}(d \psi \wedge \star d \psi) - i^{b}(i_{b} d \psi \wedge \star i_{c} d \psi)]
\ee
The term:
\be
k(1 + \alpha \psi^2) (\hat{\lambda}^{a}{}_{c} \wedge \hat{\lambda}^{c}{}_{b}) \wedge \star(e_{a} \wedge e^b \wedge e_{c})
\ee
is:
\bigskip
\be
- \frac{2 k \alpha \psi A(\psi)}{n-2}[(3-n)i_{c}(d \psi \wedge \star d \psi) -2 i^{a}[i_{a}(d \psi) \wedge \star i_{c} d \psi]]
\ee
So the non-Riemannian contribution to the Einstein $n-1$-forms $G_{c}$ is:
\bigskip
\begin{eqnarray}
-4 k \alpha \psi A(\psi) [i_{c}(d \psi \wedge \star d \psi) - i^{b}[i_{b}(d \psi) \wedge i_{c} d \psi)] - \\ \nonumber
\frac{2 k \psi A(\psi)}{n-2}[(3-n) i_{c}(d \psi \wedge \star d \psi) - 2 i^{a}[i_{a}(d \psi) \wedge \star i_{c} d \psi]]
\end{eqnarray}
\bigskip

Then the generalised Einstein equations can be written:
\bigskip
\begin{eqnarray}
k {\stackrel{o}{G}}_{c} (1 + \alpha \psi^2) - \frac{\beta'}{2}[d \psi \wedge i_{c} \star d \psi + i_{c} d \psi \wedge \star d \psi] = 0
\end{eqnarray}
In which use has been made of the relation of the relation:
\bigskip
\be
\frac{-1}{2}[A \wedge i_{a} \star A + i_{a} A \wedge \star A] = \frac{1}{2} i_{a}(A \wedge \star A) -i^{b}[i_{b}(A) \wedge \star i_{a}(A)]
\ee
\bigskip
valid for any 1-form A.\\
\bigskip
In the previous we have:
\be
\beta' = \beta + 4k \frac{n-1}{n-2}
\ee
It is interesting to note that if:
\be
\beta = -4k \frac{n-1}{n-2}
\ee
then we eliminate the kinetic term from the reduced generalised Einstein equations.\\
\\
Let us make now the comparison with the theory obtained from the action:
\be
\int \Lambda[e,\omega,\psi] = \int [k(1+ \alpha \psi^2) \stackrel{o}{R} \star 1 + \frac{\beta '}{2}(d \psi \wedge \star d\psi)]
\ee
Considering the coframe variation of the previous action we get the equation:
\be
k(1 + \alpha \psi^2){\stackrel{o}{G}}_{c} - \frac{\beta '}{2}[d\psi \wedge i_{c} \star d\psi + i_{c} d\psi \wedge \star d\psi] = 0
\ee
Which coincides with (299).\\
By using (292) we can calculate the non-Riemannian contribution to the Einstein-Hilbert term:
\begin{eqnarray}
\Delta R \star 1 & = &
\frac{1}{(n-2)^2}A^{2}(\psi) \, [i^{a}(d \psi) \wedge d \psi \wedge e_{b} \wedge \star(e_{a} \wedge e^{b}) \\ \nonumber
& - & i_{c}(d \psi) i^{c}(d \psi)(e^{a} \wedge e_{b}) \wedge \star(e_{a} \wedge e^{b})] \\ \nonumber
& + & i_{b}(d \psi) \wedge d(\psi) \wedge e^{a} \wedge \star (e^{b} \wedge e_{a}) \\ \nonumber
& = & 2 i^{a}(d \psi) \wedge d(\psi) \wedge (e_{b} \wedge \star (e_{a} \wedge e^{b})) \\ \nonumber 
& - & i_{c}(d \psi) i^{c}(d \psi) \, [(e^{a} \wedge e_{b}) \wedge \star (e_{a} \wedge e^{b})] \\ \nonumber
& = & 2 i^{a}(d \psi) \wedge d(\psi) \wedge [-g_{ba} \star e^{b} + n \star e_{a}] \\ \nonumber
& - & i_{c}(d \psi) i^{c}(d \psi) \, [e^{a} \wedge (-g_{ba} \star e^{b} + n \star e_{a})] \\ \nonumber
& = & 2(n-1) \, d(\psi) \wedge \star e_{a} \wedge i^{a}(d \psi) -n(n-1)(d \psi \wedge \star d \psi) \\ \nonumber
& = & \frac{A(\psi)^2}{(n-2)^2}(3n -2 -n^2)(d\psi \wedge \star d\psi)
\end{eqnarray}
The variation with respect to $\psi$ of (304) gives:
\be
\beta ' \, d \star d\psi - 2 k \, \alpha \, \psi \, \stackrel{o}{R} \star 1 = 0
\ee
\bigskip
\subsection{RELATION WITH A CLASS OF DILATON GRAVITY THEORIES}
\bigskip
In the previous section we saw that we can reduce the conformal generalised Einstein equations to a Levi-Civita system, before proceeding to the topic of this section let us observe that if in the reduced Einstein equations we put:
\be
\beta = 0
\ee
then we get
\be
\beta' = 4k \frac{n-1}{n-2}
\ee
which in the 4-dimensional case has the value $\beta' = 6$ which is the value required for the conformal invariance. It is clear anyway that in this case since we are fixing the metric $g_{ab} = \eta_{ab}$ the conformal transformations are not defined at least not in the usual sense.\\
\\
In this section we want to show the extension of the isomorphism to a certain class of Dilaton gravity models, consider then an action of the type:
\bigskip
\begin{eqnarray}
\int  \Lambda[e,\omega] = \int [k R \star 1 + \frac{\alpha}{2}f_{1}(\psi)(dQ \wedge \star dQ) + \frac{\beta_{0}}{2}f_{2}(\psi)(Q \wedge \star Q) \\ \nonumber
+ \frac{\beta - \beta_{0}}{2}(Q \wedge \star Q) + \frac{\delta}{2}(d \psi \wedge \star d \psi) + F[e,\omega]]
\end{eqnarray}
\bigskip
where $f_{1}(\psi)$ and $f_{2}(\psi)$ are 0-forms functions of the scalar field (dilaton), and $F[e,\omega]$ a generic $n-$form dependent on the variables $e$ and $\omega$.
Suppose that in the limit:
\bigskip
\begin{eqnarray}
f_{1}(\psi) \rightarrow 1 \\ \nonumber
f_{2}(\psi) \rightarrow 1
\end{eqnarray}
\bigskip
we get a Proca-type equation for the Weyl 1-form Q:
\bigskip
\be
\alpha d \star dQ + \beta_{0} \star Q = 0
\ee
and the theory is of Proca type, that is the generalised Einstein equations reduce to:
\be
k \, {\stackrel{o}{G}}_{c} + \tau_{c}[\alpha] + \tau_{c}[\beta_{0}] + \tau_{c}[\delta] = 0
\ee
then the isomorphism occurs in the general case with $f_{1}(\psi), f_{2}(\psi) \neq 1$ as well.\\
The proof of this result is as follows.\\
The Cartan equation can be written as:
\bigskip
\begin{eqnarray}
\alpha d(f_{1}(\psi)) \star dQ + f_{2}(\psi) \beta_{0} \star Q + (\beta-\beta_{0}) \star Q = \sum_{k}(\star E_{k}) \\ \nonumber
k D \star(e_{a} \wedge e^b) = F_{a}{}^{b}
\end{eqnarray}
in which the $E_{k}$ are a set of 1-forms which come from the Cartan equation:
\be
\underbrace{F[e,\omega]}_{\omega} = {\dot{\omega}}^{a}{}_{b} \wedge \sum_{k} \star(B_{a}{}^{b})_{k}
\ee
we have
\be
2 n \sum_{k} \star E_{k} = \sum_{k} \star (B^a{}_{a})_{k}
\ee
In the case $f_{1}(\psi) = f_{2}(\psi) = 1$ we have the equation (311) which implies:
\be
(\beta - \beta_{0}) \star Q = \sum_{k}(\star E_{k})
\ee
What happens if $f_{1}(\psi), f_{2}(\psi) \neq 1$ ? Looking back at the Cartan equation (313) we see that only the trace of the Cartan equation is modified by the replacement:
\begin{eqnarray}
\beta(Q \wedge \star Q) \rightarrow [\beta_{0} f_{2}(\psi) + (\beta - \beta_{0})](Q \wedge \star Q) \\ \nonumber
(dQ \wedge \star dQ) \rightarrow f_{1}(\psi)(dQ \wedge \star dQ)
\end{eqnarray}
The traceless part of the Cartan equation is still:
\be
k \, D \star (e_{a} \wedge e^{b}) = F_{a}{}^{b}
\ee
By using the formulas of chapter 3 we can solve for the non-Riemannian part of the connection 1-forms. Using the observation we made after formula (194) we can say that $\lambda^{a}{}_{b}$ will have the same expression as a function of $F_{a}{}^{b}$ as in the limiting case $f_{1}(\psi) = f_{2}(\psi) =1$.\\
This implies that we can calculate the quantities $E_{k}$, in particular we can say that the relation
\be
(\beta - \beta_{0}) \star Q = \sum_{k} (\star E_{k})
\ee
is valid, so that the first of (313) reduces to:
\be
\alpha \, d(f_{1}(\psi) \star dQ) + f_{2}(\psi) \, \beta_{0} \star Q = 0
\ee
Since the algebraic dependence of $\lambda^{a}{}_{b}$ on the quantities $F^{a}{}_{b}$ is the same, we can say that the following equation holds:
\be
k \, \Delta {\stackrel{o}{G}}_{c} + \tau_{c}[\beta -\beta_{0}] + \tau_{c}[F[e,\omega]] = 0
\ee
where $\tau_{c}[F[e,\omega]]$ denote the stress forms of the $F[e,\omega]$ term.\\
The generalised Einstein equations then reduce to:
\be
k \, {\stackrel{o}{G}}_{c} + f_{1}(\psi) \tau_{c}[\alpha] + f_{2}(\psi) \tau_{c}[\beta_{0}] + \tau_{c}[\delta] = 0
\ee
The extension of the isomorphism to the class of dilaton gravity theories represented by the action (309) is then proved.\\
The result obtained can be extended to actions which are more general than (309). For instance we can conceive the introduction of terms like:
\bigskip
\be
\star(F \wedge \star F) (Q \wedge \star Q) 
\ee
In which $F$ is frame independent.\\
We will get a different trace of the Cartan equation, and the non-Riemannian field equations will be isomorphic to the Levi-Civita theory in which we have a term like:
\bigskip
\be
\star(F \wedge \star F)(A \wedge \star A)
\ee
$A$ being a 1-form.\\ 
\\
Despite these changes we still have the reduction which permits us to simplify the analysis of the non-Riemannian actions.
\newpage
\section{EXACT SOLUTIONS OF NON-RIEMANNIAN GRAVITY}
\bigskip
\bigskip
In this chapter we will use the results of chapter 3 and 4 to get some exact solutions of non-Riemannian theories of gravitation.\\ 
After considering the axially symmetric Melvin solution and the time dependent Rosen solution, we will apply the results of the 4th chapter to the fascinating problem of finding black hole solutions for non-Riemannian actions. At the end of this chapter we will present a class of black-hole Dilaton solutions with non-Riemannian connection, but to begin with let us start with the static Melvin solution:\\
\\
\subsection{The axially symmetric static Melvin solution}
\bigskip
We will start the analysis of some exact solutions of non-Riemannian gravity by considering the Melvin solution, so called because it was first obtained by Melvin in the context of the Einstein-Maxwell axially symmetric solutions [30].\\
The action we are considering is very simple:\\
\be
\int \Lambda[e,\omega] = \int [k R \star 1 + \frac{\alpha}{2}(dQ \wedge \star dQ)]
\ee
with this action the connection variation gives the equation:
\be
D \star (e^a \wedge e_{b}) = 0
\ee
and the equation:
\be
d \star dQ = 0
\ee
while the coframe variation gives the Einstein Equations:
\be
k G_{a} + \frac{\alpha}{2}[dQ \wedge i_{a} \star dQ - i_{a} dQ \wedge \star dQ] = 0
\ee
by using the case 3.3.1 considered in chapter 3 we can say that the non-Riemannian part of the connection has the following very simple expression:
\be
\lambda_{ab} = -\frac{1}{2n} Q g_{ab}
\ee
and ${\hat{T}}_{c} = 0$, ${\hat{Q}}_{ab} = 0$, $T = \frac{n-1}{2n}Q$. \\
Moreover the traceless part of the non-Riemannian connection vanish:
\be
{\hat{\lambda}}_{ab} = 0
\ee
using formula (207) of the previous chapter we get that:
\be
G_{c} = {\stackrel{o}{G}}_{c}
\ee
so in this case the generalised Einstein equations coincide with the Einstein-Maxwell equations.\\ 
If we make the identification:
\be
A = Q
\ee
we can import solutions of the Einstein-Maxwell theory to the non-Riemannian case.\\
\\
We consider a cylindrical coordinate system $[t,r,\phi,z]$. \\
The metric is written as:
\be
{\bf g} = - e^0 \otimes e^0 + e^1 \otimes e^1 + e^2 \otimes e^2 + e^3 \otimes e^3
\ee
with
\begin{eqnarray}
e^0 = \sqrt{f} dt \\ \nonumber
e^1 = \sqrt{h} dr \\ \nonumber
e^2 = \frac{r}{\sqrt{f}} d \phi \\ \nonumber
e^3 = \sqrt{h} \, dz
\end{eqnarray}
Where $f$ and $h$ are functions of $r$. The non-metricity is:
\be
Q = \beta (r) \, d\phi
\ee
Where $\beta$ is a function of $r$.\\
By solving the Einstein equations we get:
\begin{eqnarray}
h = C_{1} (1 + C_{2} r^2)^2 \\ \nonumber
f = C_{3} (1 + C_{2} r^2)^2 \\ \nonumber
\end{eqnarray}
Where $C_{1}, C_{2}, C_{3}$ are constants. The solution for $\beta$ is:
\be
\beta = - \frac{1}{2}\frac{C_{4}}{(1 + C_{2} r^2)}
\ee
with $C_{4}$ another constant.\\
The constants which appear in the equations (336,337) are not unrelated but they must satisfy the constraint:
\be
C_{2} C_{3} C_{4}^2 = 16 k \alpha
\ee
\subsection{The time dependent Rosen solution}
\bigskip
In this section we will consider an axially symmetric solution which is not static, the so called Rosen solution, obtained by Rosen in his studies of the symmetries of Einstein Maxwell equations [31].\\
We start from the action:
\be
\int \Lambda[e,\omega] = \int [k R \star 1 + \frac{\alpha}{2}(dQ \wedge \star dQ)]
\ee
The variation of this action gives the same equations considered in the previous section:
\begin{eqnarray}
D \star (e^a \wedge e_{b}) = 0 \\ \nonumber
d \star dQ = 0 \\ \nonumber
k \stackrel{o}{G}_{c} + \frac{\alpha}{2}[dQ \wedge i_{c} \star dQ - i_{c}dQ \wedge \star dQ] = 0
\end{eqnarray}
where in the last equation use has been made of (331).\\
We choose as coordinates $[t,x,y,z]$. The metric tensor is chosen like the previous section with:
\begin{eqnarray}
e^0 = f_{0} dt \\ \nonumber
e^1 = f_{1} dx \\ \nonumber
e^2 = f_{2} dy \\ \nonumber
e^3 = f_{3} dz 
\end{eqnarray}
where $f_{0}, f_{1}, f_{2}, f_{3}$ are functions of $t$.\\
The non metricity 1-form $Q$ is:
\be
Q = C_{1} \, \cos t \, dx + 2C_{2} \, y \, dz 
\ee
by solving the Einstein equations we verify that:
\begin{eqnarray}
C_{1} = - 2 \, \sqrt{k} \cos \, \alpha \\ \nonumber
C_{2} = 2 \frac{\sqrt{k}}{b_{1}} \, \sin \, \alpha \\ \nonumber
f_{0} = \frac{b_{1} B(t)^{b_{2}+b_{3}}}{(\sin \, t)^2} \\ \nonumber
f_{1} = \sin \, t \\ \nonumber
f_{2} = \frac{B(t)^{b_{2}}}{\sin \, t} \\ \nonumber
f_{3} = \frac{B(t)^{b_{3}}}{\sin \, t} \\ \nonumber
b_{2} = \frac{1}{b_{3}}
\end{eqnarray}
where:
\be
B(t) =  tan[\frac{t}{2}]
\ee
We have:
\be
dQ = 2 \sqrt{k} \, \cos \, \alpha \sin \,t \,(dt \wedge dx)  + 2 \frac{\sqrt{k}}{b_{1}} \sin \, \alpha \, (dy \wedge dz)
\ee
so we can say that we have a ``gravitoelectric'' field associated with the non-metricity oscillating in the $x$ direction against a constant ``gravitomagnetic'' field in the same direction.\\
In the previous solution we have three arbitrary constants $b_{1}$, $b_{2}$ (or $b_{3}$) and $\alpha$.\\
\subsection{Non-Riemannian black hole with Dilaton}
\bigskip
In this paragraph and the next one, we want to consider the application of what we found in the previous chapter to exhibit some black-hole solutions with non-Riemannian connection and Dilaton field.\\
To begin with consider the action:
\bigskip
\be
\int \Lambda[e,\omega,\psi] = \int [k \, R \, \star 1 + \frac{\alpha}{2}e^{-2(\psi)}(dQ \wedge \star dQ) + \frac{\beta}{2}(Q \wedge \star Q) + \frac{\gamma}{2}(T \wedge \star T) + \frac{\delta}{2}(d \psi \wedge \star d \psi)]
\ee
\bigskip
The connection variation of the previous action gives:
\bigskip
\begin{eqnarray}
\alpha d(e^{-2 \psi} \star dQ) + \beta \star Q = \frac{\gamma (1-n)}{2n} \star T \\ \nonumber
k D \star (e_{a} \wedge e^b) = \delta^b{}_{a} \frac{1-n}{n} \gamma \star T + \gamma e^b \wedge i_{a} \star T
\end{eqnarray}
\bigskip
The coframe variation gives the generalised Einstein equations:
\bigskip
\be
k G_{c} + e^{-2 \psi} \tau_{c}[\alpha] + \tau_{c}[\beta] + \tau_{c}[\delta] + \tau_{c}[\gamma] = 0
\ee
\bigskip
The variation with respect to $\psi$ gives the equation:
\bigskip
\be
\delta \, d \star d \psi + \alpha  \, e^{-2 \psi}(dQ \wedge \star dQ) = 0
\ee
The second of equations (347) coincides with eq. (179). By solving it we can write the same relation (183) between $T$ and $Q$:
\bigskip
\be
T = [\frac{n-1}{2n}\frac{Q}{1 - \frac{2 \gamma (n-1)}{k n^2 (n-2)}}]
\ee
\bigskip
If we require then the constraint:
\bigskip
\be
4n^2(n-2) \beta k + (n-1)^2(n-2) \gamma k + 8 (1-n) \gamma \beta = 0
\ee
The first of equations (347) assumes the Maxwell-Dilaton form:
\be
\alpha d(e^{-2 \psi} \star dQ) = 0
\ee
As seen in the previous chapter the cancellation which occurs in the generalised Einstein equations depends on the properties of the traceless part of the Cartan equation, then we can conclude using the results of section 4.5 that the generalised Einstein equations reduce to:
\bigskip
\be
k \stackrel{o}{G}_{c} + e^{-2 \psi} \tau_{c}[\alpha] + \tau_{c}[\delta] = 0
\ee
\bigskip
We consider a spherically symmetric spacetime with metric (n=4):
\be
{\bf g} = -e^0 \otimes e^0 + e^1 \otimes e^1 + e^2 \otimes e^2 + e^3 \otimes e^3
\ee
\begin{eqnarray}
e^0 = f(r) dt \\ \nonumber
e^1 = \frac{1}{f(r)} dr \\ \nonumber
e^2 = R(r)d \, \theta \\ \nonumber
e^3 = R(r) \, \sin \, \theta \, d \phi \\ \nonumber
\end{eqnarray}
Then using the results of ref. [32-34] we can write the solution for $\psi, Q, f(r), R(r)$ as:
\begin{eqnarray}
Q = - q \, \cos \, \theta \, d\phi \\ \nonumber
\psi = - \frac{1}{2} ln(b_{1} - \frac{b_{2}}{r}) \\ \nonumber
R(r) = \sqrt{r(r-r_{1})} \\ \nonumber
f = \sqrt{1 + \frac{\alpha b_{1}q^2}{2kr_{1}r}} \\ \nonumber
\delta = - 4 k \\ \nonumber
r_{1} = \frac{b_{2}}{b_{1}}
\end{eqnarray}
\bigskip
The torsion 2-form is from (356, 184) and (188):
\bigskip
\be
T^a = \frac{8 \beta}{9 \, \gamma } \, q \, \cos \, \theta (e^a \wedge d\phi)
\ee
\bigskip
The non metricity 1-forms can be calculated using the formula:
\be
Q_{ab} = e_{a} i_{b} A_{1} + e_{b} i_{a} A_{1} - \frac{1}{2}g_{ab} A_{1} + \frac{g_{ab}}{4} Q
\ee
with:
\be
A_{1} = \frac{2 \beta q}{3 \, k} \, \cos \, \theta \, d\phi
\ee
where use has been made of (184,192,356):
Then if we satisfy the condition:
\be
\frac{\alpha b_{1}}{2 k r_{1}} < 0
\ee
we have a black hole horizon for:
\be
r = - \frac{\alpha b_{1} q^2}{2 k r_{1}}
\ee
Observe that when $ r = r_{1}$ the area with constant $r$ goes to zero.
\bigskip
\subsection{Black Hole solutions with more general actions}
\bigskip
In the previous section we considered the action (346) which is just a particular case of the class of actions which presents isomorphism with the reduced Einstein equations (353).\\
We can consider a general action which can be written like:
\bigskip
\be
\int \Lambda[e,\omega,\psi] = \int [k R \star 1 + \frac{\alpha}{2}e^{-2 \psi}(dQ \wedge \star dQ) + \frac{\beta}{2}(Q \wedge \star Q) + \frac{\delta}{2}(d \psi \wedge \star d \psi) + F[e,\omega]]
\ee
Where $F[e,\omega]$ is a term depending on the variables $e, \omega$.\\
Suppose that in the limit $e^{-2\psi} \rightarrow 1$ there is isomorphism with the theory described by the action:
\be
\int \Lambda[e,A,\psi] = \int k \stackrel{o}{R} \star 1 + \frac{\alpha}{2} (dA \wedge \star dA) + \frac{\delta}{2}(d \psi \wedge \star d \psi)
\ee
and
\be
\alpha \, d \star dA = 0
\ee
Then we can obtain black-hole solutions for the equations obtained from the action (362) according to what follows.\\
Using the results of section 4.5 we can say that if the isomorphism occurs in the limiting case $e^{-2\psi} = 1$ then the following equations have to be satisfied:
\be
\alpha \, d(e^{-2\psi} \star dQ) = 0
\ee
and
\be
k \, {\stackrel{o}{G}}_{c} + e^{-2\psi} \tau_{c} [\alpha] + \tau_{c}[\delta] = 0
\ee
where
\begin{eqnarray}
\tau_{c}[\alpha] = \frac{\alpha}{2}[dQ \wedge i_{c} \star dQ - i_{c}dQ \wedge \star dQ] \\ \nonumber
\tau_{c}[\delta] = -\frac{\alpha}{2}[d\psi \wedge i_{c} \star d\psi + i_{c}d\psi \wedge \star d\psi] \\ \nonumber
\end{eqnarray}
We can still use the ansatz (357,358), the torsion is:
\bigskip
\be
T^a = \frac{8 \beta}{9 \, \gamma} \, q \, \cos \, \theta \,(e^a \wedge d\phi)
\ee
\bigskip
The non-metricity 1-forms $Q_{ab}$ are given by the general expression (358) in which $A_{1}$ will have to be calculated using the formula:
\be
A_{1}= - (\sum_{k}A_{k} + A)
\ee
Where $A_{k}$ and $A$ are functions of the quantities $T$, $Q$, $T^c$, $\cal{Q}$, depending on which particular action is used (namely the term $F[e,\omega]$ in (362)).\\
\\
This concludes the chapter the aim of which has been to show examples of how we can use known results of Einstein-Maxwell theory and/or Dilaton theories to produce solution of a non-Riemannian theory of gravitation by means of the isomorphism properties considered in chapter 4.\\
\\
In the next Chapter we will consider the interesting problem of analysing the possible modifications of the Maxwell theory due to the introduction of a general non-Riemannian connection.
\newpage
\begin{center}
{\bf CHAPTER 6}
\end{center}
\bigskip
\section{MODIFICATION OF THE MAXWELL THEORY IN NON-RIEMANNIAN GRAVITY}
\bigskip
In this chapter we discuss the interesting problem of analysing whether the Electromagnetic theory has to be modified in a non-Riemmannian space time. In a recent paper by Vandyck [35] a claim has been given about the non equivalence of the Electromagnetic theory in the formulation which uses the exterior differentiation and the one which uses covariant differentiation.\\
This statement is an artificial one in the sense it deliberately ignores the fundamental principles of electromagnetism, like the conservation of electric and magnetic charge.\\
If we start from these postulates there is one and only one possible Maxwell theory as shown by Hehl and others [36]. The only possible freedom is in the so called constitutive law which being dependent on the metric is somehow affected by possible coupling to non-Riemannian fields. After presenting how the conservation axioms do allow us to formulate a unique Maxwell theory, we use the formalism of chapter 2 and 3 to show some cases of modification of the constitutive law.
\newpage
\subsection{Conservation axioms and Maxwell theory}
\bigskip
Consider a theory in which we have the electric current $3$-form $J$. We assume that a $(1+3)$ foliation of spacetime holds locally. We can label the different three dimensional hypersurfaces by the parameter $\tau$.\\
We may write the decomposition of the current $3$-form as:
\be
J = \rho - j \wedge d \tau
\ee
where $\rho$ is the charge density $3$-form and $j$ the electric current $2$-form.\\
The first axiom we assume is the conservation of the electric charge:
\be
\oint_{\partial V}J = \int_{V} d J = 0
\ee
with $V$ an arbitrary volume and $ \partial V$ its boundary.\\
If (371) is valid for any $\partial $ then $J$ is closed and hence locally:
\be
dG = J
\ee
\bigskip
The second axiom concerns the definition of the electromagnetic field strength $F$ by the Lorentz force expression:
\be
f_{a} = i_{a}F \wedge J
\ee
where $f_{a}$ is the force-density 4-form.\\
The $2$-form $F$ can be decomposed as:
\be
F = B + E \wedge d \tau
\ee
The Lorentz force expression yields an operational definition of the Electromagnetic field $F$ as a force field.\\
In the previous definition no reference has been made to metric or connections.\\
We need now another postulate which is related to the conservation of the magnetic charge which can be written as:
\be
\oint_{\partial V_{3}}F = \int_{V_{3}} dF = 0
\ee
from which we get:
\be
dF = 0
\ee
If we introduce a normal vector $n$ such that $i_{n} d \tau = 1$, we can then introduce the three dimensional exterior derivative:
\be
\b{d} = d - d\tau \, i_{n} \wedge
\ee
so that we can write (376) as:
\begin{eqnarray}
{\b{d}}B = 0 \\ \nonumber
{\b{d}}{E} + \dot{B} = 0
\end{eqnarray}
Equations (372) and (376) are the Maxwell equations.\\
In this approach the general covariance is manifest, moreover the introduction of a metric and a connection does not affect the structure of the previous equations.\\
Therefore in this framework the Maxwell equations are the same in a Minkowskian, Riemannian or non-Riemannian spacetime. Any space time geometry is irrelevant, in particular neither torsion nor non-metricity modifies the structure.\\
In spite of deformation of spacetime by means of gravity, the Maxwell equations remain 'stable'.\\
\\
In his paper Vandyck claims that different theories may be obtained whether we use the covariant approach or the exterior one.\\
In the covariant approach we write the Maxwell equations in the form:
\begin{eqnarray*}
{\nabla}^{a}F_{ab} = -4 \pi j_{b} \\ \nonumber
{\nabla}[_{a} \, F_{bc}] = 0 \\ \nonumber
\end{eqnarray*}
where the square bracket indicates the antisymmetrization with respect to the indices $a,b,c$.\\
It is clear that the general covariant derivative will contain certain contributions dependent on the torsion and the non-metricity. Those terms will in general violate the conservation of electric and magnetic charge.\\
This violation does not occur if we start from equations (372, 376) since the exterior derivative may be defined independently on the connection.\\
Equations (372) and (376) for $G$ and $F$ are un undetermined system of equations.\\
It is necessary to introduce  another relation which relates $F$ and $G$ in order to reduce the number of independent variables:
\be
G = G(F)
\ee
this relation is called the {\bf Constitutive law}.\\
While in the Riemannian case $G = \star F$ in vacuo, we may contemplate modifications to this relation in the presence of non-Riemannian fields.\\
In the following we present two cases of modifications, the first one in which the modification of the constitutive law is due to the torsion of the spacetime, the second in which the non-metricity plays a role.\\
\bigskip
\subsection{Field equations with the term $\star(T^{c} \wedge e_{c}) \wedge F \wedge A$}
\bigskip
Consider the theory obtained from the action ($n=4$):
\be
S = \int [ k \, R \star 1 + \frac{\alpha}{2}(F \wedge \star F) - \beta \star(T^{a} \wedge e_{a}) \wedge (F \wedge A)]
\ee
where $T^{a}$ denote the torsion two forms, $A$ is the electromagnetic potential defined by:
\be
F = dA
\ee
If we define the axion field with the expression:
\be
d \theta = \star(T^{a} \wedge e_{a})
\ee
we have
\be
d[\theta (F \wedge A)] = d\theta \wedge F \wedge A + \theta(F \wedge F)
\ee
so that mod d:
\be
d\theta \wedge F \wedge A = - \theta (F \wedge F)
\ee
this means that the theory described by the action (380) is equivalent to the theory described by the action:
\be
S = \int [k \, R \star 1 + \frac{\alpha}{2}(F \wedge \star F) + \beta \, \theta(F \wedge F)
\ee
this action in the literature is known as the 'Axion theory' [37-39].\\
\\
By considering the connection variation of the action (380) we get:
\be
k D \star (e_{a} \wedge e^{b}) = \beta (e^{b} \wedge e_{a}) \wedge \star (F \wedge A)
\ee
The coframe variation instead gives the generalised Einstein equations:
\begin{eqnarray}
k \, G_{c} + \frac{\alpha}{2}[F \wedge i_{c} \star F - i_{c}F \wedge \star F] \\ \nonumber
- \beta[i_{c}(F \wedge A) \wedge \star(T^{a} \wedge e_{a}) + (T^{a} \wedge e_{a}) \wedge i_{c} \star(F \wedge A) \\ \nonumber
+ \lambda^{a}{}_{c} \wedge e_{a} \wedge \star (F \wedge A) + T_{c} \wedge \star (F \wedge A)] = 0
\end{eqnarray}
The variation with respect to $A$ gives:
\be
\alpha \, d\star F + \beta \star (T^{a} \wedge e_{a}) \wedge F = 0
\ee
Let us now solve the Cartan equation for the non-metricity and torsion.\\
We have:
\be
f^{cb}{}_{a} =\frac{\beta}{k} \, i^{c} \star [(e^{b} \wedge e_{a}) \wedge \star (F \wedge A)]
\ee
In the following we define $ \gamma = \frac{\beta}{k}$.\\
From the symmetry property $f_{cba} = - f_{cab}$ we get using formula (90) in chapter 3:
\be
{\hat{Q}}_{ab} = 0
\ee
where ${\hat{Q}}_{ab}$ denotes the traceless part of the non-metricity 1-forms.\\
From the expression of $f_{cab}$ we get:
\be
f^{ca}{}_{c} = 0
\ee
then using relation (91) we find:
\be
T = \frac{3}{8}Q
\ee
To calculate the traceless part of the torsion we use formula (92) which in our case becomes:
\be
{\hat{T}}_{c} = \frac{1}{3}(e_{c} \wedge e^{a}) f^{d}{}_{ad} + \frac{1}{2}(e^{b} \wedge e^{a}) f_{cba}
\ee
Using (389) we obtain:
\be
{\hat{T}}_{c} = \gamma i_{c}(F \wedge A)
\ee
so that
\be
T^{c} = \gamma i^{c}(F \wedge A) + \frac{1}{3}(e^{c} \wedge T)
\ee
from which we get:
\be
\star(T^{a} \wedge e_{a}) = 3 \gamma \star (F \wedge A)
\ee
so that the modified Maxwell equation can be written as:
\be
\alpha  \, d \star F + 3 \beta \, \gamma \, \star (F \wedge A) \wedge F = 0
\ee
The solution for the non-metricity and torsion can be written as:
\begin{eqnarray}
Q_{ab} = \frac{1}{4}g_{ab} Q \\ \nonumber
T^{a} = \gamma i^{a}(F \wedge A) + \frac{1}{3}(e^{a} \wedge T) \\ \nonumber
T = \frac{3}{8}Q
\end{eqnarray}
Equations (376) and (397) give the modified Maxwell theory.\\
Let us calculate now the non-Riemannian part of the connection 1-forms $\lambda^{a}{}_{b}$.\\
By using formula (23) and (398) and (392) we find:
\be
\lambda^{a}{}_{b} = \frac{3}{2} \, i^{a}i_{b}(F \wedge A) - \frac{1}{8} \delta^{a}{}_{b} Q
\ee
and its traceless part:
\be
{\hat{\lambda}}^{a}{}_{b} = \frac{3}{2}i^{a}i_{b}(F \wedge A)
\ee
By using this expression we can calculate the non-Riemannian contribution to the Einstein 3-form $\Delta G_{c}$:
\begin{eqnarray}
\Delta G_{c} = {\hat{\lambda}}^{a}{}_{d} \wedge {\hat{\lambda}}^{d}{}_{b} \wedge \star(e_{a} \wedge e^{b} \wedge e_{c}) = \\ \nonumber
\frac{9}{4}[i^{a}i_{d}(F \wedge A) \wedge i^{d}i_{b}(F \wedge A) \wedge \star (e_{a} \wedge e^{b} \wedge e_{c})]
\end{eqnarray}
\\
\subsection{Field equations with the term $ \star (Q \wedge \star Q)(F \wedge F) $}
\bigskip
In this section we want to consider another case of modification of the constitutive law due to the introduction in the action of the term $ \star(Q \wedge \star Q)(F \wedge F)$.\\
Consider the theory obtained from the action:
\be
S = \int [k \, R \star 1 + \frac{\alpha}{2}(F \wedge \star F) + \frac{\gamma}{2}(dQ \wedge \star dQ) + \frac{\beta}{2} \star (Q \wedge \star Q) (F \wedge F)]
\ee
The term $\star (Q \wedge \star Q)(F \wedge F) $ can be written:
\be
\theta (F \wedge F)
\ee
with
\be
\theta = \star(Q \wedge \star Q)
\ee
The coframe variation gives the Einstein equation:
\begin{eqnarray}
k \, G_{c} + \frac{\alpha}{2}[F \wedge i_{c} \star F - i_{c}F \wedge \star F] \\ \nonumber
+ \frac{\gamma}{2}[dQ \wedge i_{c} \star dQ - i_{c}dQ \wedge \star dQ] - \\ \nonumber
\frac{\beta}{2}[i_{c}(F \wedge F) \wedge \star (Q \wedge \star Q) + (Q \wedge i_{c} \star Q + i_{c}Q \wedge \star Q) \wedge \star (F \wedge F)] = 0
\end{eqnarray}
The connection variation gives the Cartan equation:
\be
kD \star (e_{a} \wedge e^{b}) = 2 \delta^{b}{}_{a} \, (\gamma \, d \star dQ + \beta \star Q \star (F \wedge F))
\ee
which is equivalent to the equations:
\begin{eqnarray}
k \, D \star (e_{a} \wedge e^{b}) = 0 \\ \nonumber
\gamma \, d \star dQ + \beta \star Q \star (F \wedge F) = 0
\end{eqnarray}
The variation with respect to $A$ gives the equation:
\be
\alpha \, d \star F + \beta \, d[\star (Q \wedge \star Q)] \wedge F = 0
\ee
Solving the first of equations (407) with respect to the non-metricity and torsion we have:
\begin{eqnarray}
Q_{ab} = \frac{1}{4}g_{ab}Q \\ \nonumber
T^{a} = \frac{1}{3} (e^{a} \wedge T)\\ \nonumber
T = \frac{3}{8}Q
\end{eqnarray}
The non-Riemannian part of the connection is:
\be
\lambda_{ab} = - \frac{1}{8}g_{ab} Q
\ee
using the previous relation we see that:
\be
G_{c} = {\stackrel{o}{G}}_{c}
\ee
From the second of equations (407) we can see that the Weyl field $Q$ acquires mass due to the dynamical term $\star(F \wedge F)$. \\
\\
In conclusion only the constitutive law can be modified by post-Riemannian structures. We presented two examples of such modifications.\\
\newpage
\section{CONCLUDING REMARKS}
\bigskip
In this thesis the remarkable property of cancellation which exists in the field equations of certain non-Riemannian models of gravitation has been revisited in order to generalise what obtained by Tucker, Wang and others few years ago [22,25,26].\\
In this work the relation between non-Riemannian gravity and General Relativity is obtained presenting a step by step proof by hand valid for any dimension of the space time. The insight obtained by this proof is then used to extend the simplification property to certain models of scalar theories of gravity. The extended version of the theorem is applied to get some new black hole solutions with non-Riemannian connection as well as some solutions obtained simply importing the know solutions from the Riemannian-Einstein-Maxwell case.\\
The step by step proof has been obtained by systematically studying certain particular solutions of the Cartan equation. This particular solutions have been used to obtain the relation for the non-Riemannian part of the connection which has been used in the 4th chapter to verify the existence of the cancellation property in the field equations.\\
A particular role in the models considered in this thesis is assumed by the Weyl form $Q$. It satisfies a differential equation which for the so called 'Proca-type' theories is of the form:
\be
\alpha \, d\star dQ + \beta \star Q = 0
\ee
Both the torsion two forms and the non-metricity 1-forms can be expressed in term of $Q$ and certain algebraic expressions, containing contractions and products of coframe vectors.\\
As stated a fundamental role in this work is assumed by the Cartan Equation.\\
From the Cartan equation we extract:\\
\\
1] The trace part which gives the above mentioned equation for $Q$.\\
\\
2] The traceless part which gives the non-Riemannian part of the connection $\lambda^{a}{}_{b}$  as a function of $Q$ and the coframe vectors.\\
\\
The observation that the non-Riemannian part of the connection is given by the traceless part of the Cartan equation is significant. It means that if we introduce a 'diagonal' modification \footnote{We mean terms in the action whose connection variation can be written as ${\dot{\omega}}^{a}{}_{b} \wedge {\delta}^{b}{}_{a} F$ where $F$ is a $n-1$ form} in the action like those represented by the expression (309), though we may modify the equation for $Q$ the algebraic expression for $\lambda^{a}{}_{b}$ as a function of $Q$ is still the same.\\
This has far reaching consequences because the simplification which occurs in the generalised Einstein equations depends on the algebraic  expression of $\lambda^{a}{}_{b}$ as a function of $Q$.\\
We have shown how this can be used to extend the isomorphism to other models of non-Riemannian gravity and applied to the task of finding black-hole Dilaton solutions with non-Riemannian connections.\\
 We have considered the effect of post-Riemannian structures on the Maxwell theory. The recent claim by Vandyck of different Maxwell theories depending on whether exterior or covariant derivative approach is used, appears to be incorrect, since as shown by Hehl and others, if we do accept that the conservation laws are the base of the electromagnetic theory then only one theory is possible.\\
Only the constitutive law being related to the metric can be affected by post-Riemannian structures of the space time. Using the results of chapter 2 and 3, two cases have been considered of such a modification.\\
In appendix B we have obtained the expression of the autoparallels for the case of the Proca-type theories.\\
In conclusion the achievements of the present work have been:\\
\\
$\bullet$ \emph{A systematic study and a classification of certain solutions of the Cartan equation}.\\
\\
$\bullet$ \emph{The extension of the cancellation property of the field equations of the Proca type models of non-Riemannian gravity to the general $n$ dimensional case}.\\
\\
$\bullet$ \emph{The study of the generalised field equations in the conformal case}.\\
\\
$\bullet$ \emph{The proof of the cancellation property for a class of Scalar theories of gravity}.\\
\\
$\bullet$ \emph{New solutions to the generalised field equations, in particular a class of black-hole solutions with non-Riemannian connections}.\\
\\
$\bullet$ \emph{The study of some examples of modification of the constitutive law in electromagnetic theory in the non-Riemannian case}.\\
\\
$\bullet$ \emph{The calculation of the autoparallels in the Proca-type case}.\\
\newpage
\begin{center}
\section{\bf APPENDICES}
\end{center}
\bigskip
\bigskip
\begin{center}
{\bf APPENDIX A}
\end{center}
\bigskip
\begin{center}
{\Large\sc NO-HAIR THEOREM FOR}
\\[5pt]
{\Large\sc SPHERICAL SCALAR BLACK HOLES}
\end{center}
\bigskip
\subsection{Riemannian case}
\bigskip
In this appendix we want to outline a result concerning the impossibility of finding black-hole solutions with scalar fields in a spherical symmetric case, we will outline the Riemannian case first, then we will discuss why the theorem recently formulated by Banerjee and Sen [40] in general is not applicable in non-Riemannian gravity.\\
We will follow the notation used in Ref [40]. Consider a general scalar tensor theory of gravity coupled  to a Maxwell field:
\be
S[g_{\mu \nu}, \psi, F_{\mu \nu}]  = \int [f(\psi) R - h(\psi) g^{\mu \nu} \psi_{, \mu} \psi_{, \nu} - F_{\mu \nu}F^{\mu \nu}] \sqrt{-g} d^{4} x
\ee
where $g_{\mu \nu}, \psi$ and $F_{\mu \nu}$ are the metric tensor, the scalar field, and the Maxwell tensor field respectively ($n=4$).\\
As a result of the non-minimal coupling of the scalar field to gravity the Newtonian constant $G$ becomes a function of $\psi$. The previous action reduces to Brans-Dicke theory [41] for $f(\psi) = \psi$ and $h(\psi) = \frac{\omega}{\psi}$ with $\omega$ constant parameter. When $f(\psi) = 1 - \frac{1}{6}\psi^2$ and $h(\psi) = \frac{1}{2}$ we get the conformally coupled scalar field of Bekenstein [42].\\
The study of the general action (413) is in general quite difficult but is possible to reduce the study of (413) to a simpler action by considering the conformal mapping:
\be
g_{\mu \nu} \rightarrow \Omega^2 \, g_{\mu \nu} 
\ee
that is by defining the new metric:
\be
{\z{g}}_{\mu\nu} = \Omega^2 g_{\mu\nu}
\ee
The function $\Omega^2$ has to be chosen as:
\be
\Omega^2 = f(\psi)
\ee
The Ricci scalar transforms as:
\be
\z{R} =\Omega^2 R + 6 \Omega^{-1} \Delta \Omega
\ee
The scalar field is redefined as:
\be
\z{\psi}(\psi) = \sqrt{2} \int_{\psi_{0}}^\psi \, d\xi \, \sqrt{\frac{3}{2}(\frac{d}{d \xi}ln \, f(\xi))^2 + \frac{h(\xi)}{f(\xi)}}
\ee
so that the action (413) reduces to:
\be
\z{S}[{\z{g}}_{\mu\nu},\z{\psi},{\z{F}}_{\mu\nu}] = \int [\z{R} - \frac{1}{2}\z{g}_{\mu\nu} {\z{\psi}}_{,\mu} {\z{\psi}}_{,\nu} - {\z{F}}^2]\sqrt{-\z{g}}d^4 x
\ee
where $F^2 = F_{\mu\nu}F^{\mu\nu}$. \\
We have used the fact for ($n=4$)
\be
\sqrt{-g}F^2 = \sqrt{- \z{g}}{\z{F}}^2
\ee
The field equations of action (413) become:
\begin{eqnarray}
{\z{G}}_{\mu\nu} = {\z{\psi}}_{, \mu}{\z{\psi}}_{, \nu} - \frac{1}{2} {\z{g}}_{\mu\nu} \, \z{\psi}^{, \alpha} {\z{\psi}}_{, \alpha} - {\z{g}}^{\alpha \beta} {\z{F}}_{\alpha\mu}{\z{F}}_{\alpha\nu} + \frac{1}{4} {\z{G}}_{\mu\nu} {\z{F}}_{\alpha\beta} {\z{F}}^{\alpha \beta} \\ \nonumber
({\z{F}}^{\mu\nu})_{; \nu} = 0 \\ \nonumber
\Delta \z{\psi} = 0
\end{eqnarray}
The solution for this set of equations had been given by Penney [27] . In the case of a static spherically symmetric spacetime the line element can be written as:
\be
ds^2 = - e^{\z{\gamma}} dt^2 + e^{\z{\alpha}}dr^2 + e^{\z{\beta}} (d \theta^2 + {\sin}^2 \theta d \phi^2)
\ee
We can choose a coordinate system in which:
\be
\z{\alpha} + \z{\gamma} = 0
\ee
In such a way the solution is:
\begin{eqnarray}
e^{\z{\alpha}} = e^{-\z{\gamma}} = (r-a)^{-\Lambda}(r-b)^{-\Lambda}[\frac{b(r-a)^{\Lambda} - a(r-b)^{\Lambda}}{(b-a)}]^2 \\ \nonumber
e^{\z{\beta}} = (r-a)^{1- \Lambda}(r-b)^{1- \Lambda}[\frac{b(r-a)^{\Lambda} - a(r-b)^{\Lambda}}{(b-a)}]^2 \\ \nonumber
\z{\psi} \, = \frac{c}{a-b} ln \frac{r-a}{r-b} = \sqrt{\frac{1-\Lambda^{2}}{2}} ln( \frac{r-a}{r-b})
\end{eqnarray}
and
\be
{\z{F}}_{14} = q e^{-{\z{\beta}}}
\ee
The constants $\Lambda, a, b, c$ are related by the equations:
\begin{eqnarray}
2\Lambda^2 ab = q^2 \\ \nonumber
a + b = 2m \\ \nonumber
\Lambda^2 c^2 = (1-\Lambda^2)(2 \Lambda^2 m^2 - q^2)
\end{eqnarray}
$q$ and $m$ are recognised to be the total charge and mass of the static sphere, $c$ is an integration constant related to the scalar charge.
The constant $\Lambda$ has to take values between $0$ and $1$ ( to have a real scalar field).\\
For $\Lambda = 1, c=0$ and the scalar field becomes trivial and the metric goes into the usual Reissner-Nordstrom (RN) solution [43-44].\\
The curvature scalar for the line element (422) is calculated to be:
\bigskip
\be
R = \frac{(a-b)^2}{2(r-a)^{2-\Lambda}(r-b)^{2-\Lambda}}\frac{(1-\Lambda^2)}{[b(r-a)^{\Lambda} - a(r-b)^{\Lambda}]^2}
\ee
\bigskip
From this expression we can see that the solution is asymptotically flat. We can see that the metric components are singular for $r=a$ and $r=b$, moreover the previous expression for $R$ shows that the scalar curvature blows up if $\Lambda \neq 1$, so we get physical singularities.\\
On the other hand if $\Lambda = 1$ the scalar field becomes trivial and for these surfaces we get:
\be
R = 0
\ee
and the line element reduces to the Reissner-Nordstrom one.\\
The conclusion is that if we have a scalar field the only possible solution in a spherically symmetric space-time is the RN black-hole.\\
Using the mappings (415) and (416) we can move from the action (419) to the more general one (413), the conclusion is that in a spherically symmetric space-time it is NOT possible to have a non trivial scalar field.\\
\\
\subsection{Non-Riemannian case}
\bigskip
Can we generalise this result to non-Riemannian gravity ?\\
The answer in general is no. Because if we use relation (34) we can write the scalar curvature in general as:
\be
R = \stackrel{o}{R} + \, i_{a}i_{c} (\stackrel{o}{D} {\hat{\lambda}}^{ca} + {\hat{\lambda}}^{c}{}_{d} \wedge {\hat{\lambda}}^{da})
\ee
so in general in the total curvature we have extra terms which are not affected by the metric scaling so that the proof of the previous paragraph cannot be extended to non-Riemannian gravity.\\
The situation is even worse if we try to apply the isomorphism between Einstein-Maxwell system and non-Riemannian theories to the case in which $F \wedge \star F$ is replaced by:
\be
dQ \wedge \star dQ
\ee
because in general if we perform the transformation $g_{\mu\nu} \rightarrow \Omega^2 g_{\mu\nu}$ we get since $Q_{ab} = Dg_{ab}$
\be
Q_{ab} \rightarrow d(\Omega^2)g_{ab} + \Omega^2 Q_{ab}
\ee
or
\be
Q \rightarrow n d(\Omega^2) + \Omega^2 Q
\ee
so that:
\be
dQ \rightarrow d(\Omega^2)Q + \Omega^2 \, dQ
\ee
That would mean the appearance of Proca related terms in the action which are in general very difficult to treat.\\
So the problem of establishing a theorem (if any) of scalar no-hair for non-Riemannian gravity is not simple, and further investigations will have to address these questions.\\
Let us observe that the theorem formulated in section 8.1 cannot be applied to what was considered in sections 4.5, 5.3, 5.4, because in those cases we have the Dilaton field directly coupled to the Electromagnetic field and/or to $dQ$ through terms like $h(\psi)(F \wedge \star F)$.\\
In general, these cases which occur in Dilaton gravity escape the no-hair theorem, since it is not possible to remove the couplings between the scalar field $\psi$ and $F$ by using the transformations (415,418).\\
\newpage
\begin{center}
{\bf APPENDIX B}
\end{center}
\bigskip
\begin{center}
{\Large\sc AUTOPARALLELS IN NON-RIEMANNIAN}
\\[5pt]
{\Large\sc GRAVITY}
\end{center}
\bigskip
\subsection{Autoparallels in non-Riemannian gravity}
\bigskip
In this appendix we want to obtain a general expression for the autoparallels in non-Riemannian gravity.\\
Consider a generic curve $C$ with tangent vector $\dot{C}$, we say that it is autoparallels (of $\nabla$) if it satisfies:
\be
{{\nabla}_{\dot{C}}} \dot{C} = 0
\ee
This condition coincides with ${\stackrel{o}{\nabla}}_{\dot{C}} \dot{C} = 0$ in Riemannian gravity, but in general if the connection is non-Riemannian the two expressions will differ.\\
From relation (14) and (16) we get:
\begin{eqnarray}
{\nabla}_{\dot{C}} \dot{C} - {\stackrel{o}{\nabla}}_{\dot{C}} \dot{C} = \lambda(\dot{C}, \dot{C} ,-) \\ \nonumber
\lambda^{a}{}_{b}(\dot{C}) = \lambda(\dot{C}, X_{b}, e^{a})
\end{eqnarray}
We write the tangent vector as:
\be
\dot{C} = {\dot{C}}^{a}X_{a}
\ee
where $X_{a}$ are a set of frame vectors.\\
The condition (434) gives:
\be
e^{a}({\stackrel{o}{\nabla}}_{\dot{C}} \dot{C}) = -{\lambda^{a}}{}_{d}(\dot{C}) {\dot{C}}^{d}
\ee
Using the expression (23) we find:
\be
e_{a}[{\stackrel{o}{\nabla}}_{\dot{C}} \dot{C}] = -\frac{{\dot{C}}^{d}}{2}[i_{a}T_{d} -i_{a}T_{a} -i_{a}i_{b}T_{c}e^{c} - (i_{d}Q_{ac} - i_{a}Q_{dc})e^c - Q_{ad}]
\ee
or
\be
{\stackrel{o}{\nabla}}_{\dot{C}} \dot{C} = -\frac{{\dot{C}}^{d}}{2}[i_{a}T_{d} -i_{a}T_{a} -i_{a}i_{b}T_{c}e^{c} - (i_{d}Q_{ac} - i_{a}Q_{dc})e^c - Q_{ad}]X^{a}
\ee
So as a result of the non-Riemannian connection 'forces' appear related to the torsion and the non-metricity of the spacetime.\\
In the next section we will consider the autoparallels in the 'Proca-type' case.\\
\\
\subsection{Autoparallels in the Proca-type case}
\bigskip
We want now to calculate the autoparallels in the case in which we consider Proca-type theories.\\
In this case we can use the expression (189) for the non-Riemannian part of the connection
\be
\lambda^{a}{}_{b} = \frac{1}{2n}[e^{a} \wedge i_{b}(A_{3} + 2A_{1} - A_{2}) + e_{b} \wedge i^{a}(A_{2} -(2+2n)A_{1} -A_{3})+\delta^{a}{}_{b}(2A_{1} - A_{2})]
\ee
where
\begin{eqnarray}
A_{3} = \frac{n-1}{2n}T \\ \nonumber
A_{2} = Q \\ \nonumber
A_{1} = -(A + \sum_{k}A_{k})
\end{eqnarray}
Substituting expression (440) in relation (437) we get after a simple calculation:
\be
{\stackrel{o}{\nabla}}_{\dot{C}} \dot{C} = \frac{1}{2n}[\dot{C}(A_{3} \dot{C}) + 4 \dot{C}(A_{1} \dot{C}) - 2 \dot{C}(A_{2} \dot{C}) + (\dot{C})^2 (\z{A_{2}} - (2+2n) \z{A_{1}} - \z{A_{3}})]
\ee
where the products like $A_{3} \dot{C}$ indicate $A_{3}(\dot{C})$. $\z{A}$ is a vector defined by:
\be
\z{A} = A^a X_{a}
\ee
for a generic 1-form $A = A^{a}e_{a}$.\\
Let us consider the application of relation (442) to the case considered in section 4.1, in that case we have:
\begin{eqnarray}
A_{1} = \frac{\gamma}{n \, k}T \\ \nonumber
A_{2} = Q = \frac{\gamma(1-n)}{2 \, \beta \, n}T \\ \nonumber
A_{3} = \frac{n-1}{2n}T
\end{eqnarray}
so that:
\be
{\stackrel{o}{\nabla}}_{\dot{C}} \dot{C} = \frac{1}{2n} [\dot{C}(A_{4} \dot{C}) + (\dot{C})^2 A_{5}]
\ee
where
\begin{eqnarray}
A_{4} = A_{3} + 4A_{1} - 2A_{2} = \frac{T}{2nk \beta}[(n-1)k \beta + 8 \gamma \beta - 2 \gamma k (1-n)]\\ \nonumber
A_{5} = \frac{\z{T}}{2nk\beta}[\gamma k (1-n) - 4(1+n) \gamma \beta - k \beta (n-1)]
\end{eqnarray}
We can ask if by choosing in a proper way the constants we can satisfy the condition:
\be
{\stackrel{o}{\nabla}}_{\dot{C}} \dot{C} = 0
\ee
for a generic curve $C$, that is we can eliminate the non-Riemannian forces, to this aim we need:
\begin{eqnarray}
[(n-1)k\beta + 8 \gamma\beta - 2\gamma k (1-n)] = 0 \\ \nonumber
[\gamma k(1-n) - 4(1+n)\gamma \beta - k \beta(n-1)] = 0
\end{eqnarray}
The only non trivial solution of this system is given by:
\begin{eqnarray}
k = k\\ \nonumber
\beta = \frac{k}{4} \\ \nonumber
\gamma = -\frac{1}{8}\frac{k(n-1)}{n} \\ \nonumber
\end{eqnarray}
It is easy to see that this solution does not satisfy the condition (186).\\
The case of massive Weyl field is quite analogous with $\beta$ being replaced by $\beta -\beta_{0}$ and (186) being replaced by (224).\\
So we conclude  that it is not possible by a suitable choice of the constants to satisfy the condition ${\stackrel{o}{\nabla}}_{\dot{C}} \dot{C} = 0$ for any curve $C$. 
\newpage
\begin{center}
{\bf APPENDIX C}
\end{center}
\bigskip
\begin{center}
{\Large\sc DIFFERENTIAL GEOMETRY AND}
\\[5pt]
{\Large\sc EXTERIOR CALCULUS}
\end{center}
\bigskip
\subsection{Differential Geometry and exterior calculus}
In this appendix we review some basic definitions of differential geometry in the exterior calculus formulation.\\
We start from the concept of metric; in the usual formalism we write the interval in the form:
\be
ds^2 = g_{\mu \nu} dx^{\mu} dx^{\nu}
\ee
The metric tensor {\bf g} is defined by:
\be
g = g_{\mu \nu} dx^{\mu} \otimes dx^{\nu}
\ee
If the metric is diagonal we have:
\be
g = g_{\mu \mu} dx^{\mu} \otimes dx^{\mu} 
\ee
which can be rewritten as:
\be
g = e^{a} \otimes e_{a}
\ee
where the coframe is defined as:
\be
e^{a} = \sqrt{g_{aa}} dx^{a}
\ee
The dual frame vectors $X_{b}$ are defined by:
\be
e^{a}(X_{b}) = \delta^{a}{}_{b}
\ee
The previous relation defines an orthonormal frame system. If we consider such a frame the calculations may be greatly simplified.\\
One the most used concepts is the one of exterior multiplication $\wedge$. To define the wedge product consider first its effect on the coframe vectors:
\be
e^{\alpha_{1}} \wedge e^{\alpha_{2}} \wedge ..... \wedge e^{\alpha_{n}}
\ee
If we consider a generic permutation of the above I get the same expression multiplied by a factor $(-1)^p$ where $p$ indicates the number of transpositions with respect to the permutation $\alpha_{1}, \alpha_{2}, ..... \alpha_{n}$.\\
A generic $p$ form is defined by:
\be
\psi = \frac{1}{p!} \, \psi_{\beta_{1} \beta_{2} ...\beta_{p}} \, e^{\beta_{1}} \wedge e^{\beta_{2}} \wedge .....\wedge e^{\beta_{p}}
\ee
The exterior multiplication satisfies the distributive and the associative property and for two generic forms we have:
\be
\Phi \wedge \Pi = (-1)^{pq}(\Pi \wedge \Phi)
\ee
where $\Phi$ and $\Pi$ are a $q$ form and a $p$ form respectively.\\
By using the duality $e^{a}(X_{b}) = \delta^{a}{}_{b}$ we can define the contraction operator $i_{b}$ also called interior multiplication:
\be
i_{b}e^{a} = \delta^{a}{}_{b}
\ee
which satisfies the property:
\be
i_{b}(\Phi \wedge \Psi) = i_{b}\Phi \wedge \Psi + (-1)^{p} \Phi \wedge i_{b} \Psi
\ee
where $\Phi$ and $\Psi$ are a $p$ and a $q$ form respectively.\\
The following properties follow from the definition:
\begin{eqnarray}
i_{a}i_{b} A = -i_{b}i_{a} A \\ \nonumber
p \, A = e^{a} \wedge i_{a} A 
\end{eqnarray}
for any generic $p$ form $A$.\\
\\
The exterior derivative $d$ maps a $p$-form into a $(p+1)$ form. Its effects is basically to calculate the derivatives with respect to the variables and then wedge multiplying by the differentials of the coordinate. For example considering a generic function $f(x,y,z)$ we have:
\be
df = dx \wedge f_{x}(x,y,z) + dy \wedge f_{y}(x,y,z) + dz \wedge f_{z}(x,y,z)
\ee
Analogously for any form $p$-form $\alpha$.\\
From the definition it follows that:
\begin{eqnarray}
d(d(\Psi)) = 0 \\ \nonumber
d(\alpha \wedge \beta) = d \alpha \wedge \beta + (-1)^p \alpha \wedge d \beta
\end{eqnarray}
where $p$ is the degree of $\alpha$.\\
\\
The Hodge star operation $*$ associated with a generic metric is defined by its effect on the coframe vectors as:
\be
\star(e^1 \wedge e^2 \wedge .... \wedge e^p) = i^{p}i^{p-1} .....i^{1} \star 1
\ee
where $\star1$ is the volume form $\omega = e^1 \wedge e^2 \wedge ..... \wedge e^n$. It satisfies the property:
\begin{eqnarray}
\alpha \wedge \star \beta = \beta \wedge \star \alpha
\end{eqnarray}
for any two $p$-forms $\alpha$ and $\beta$.\\
The metric dual of a vector $\stackrel{-}{X}$ is defined by:
\be
\stackrel{-}{X}(Y) = g(X,Y)
\ee
It then follows the important relation:
\be
\star(\phi \wedge e_{a}) = i_{a} \star \phi
\ee
for any form $\phi$.\\ 
\\
Another important object which is very often used to simplify the notations in the frame independent approach is the covariant exterior derivative, in order to define that, we need the notion of linear connection. If we consider an arbitrary local basis of vector fields the most general connection can be specified by writing:
\be
{\nabla_{X_{a}}}X_{b} = \Lambda^{c}{}_{b}(X_{a})X_{c}
\ee
where $\Lambda^{c}{}_{b}$ are a set of $n^2$ 1-forms in an $n$ dimensional manifold. The previous one can be rewritten as:
\be
{\nabla_{X_{a}}}X_{b} = \Gamma^{c}{}_{ab} X_{c}
\ee
Where $\Gamma^{c}{}_{ab}$ are the generalised Christoffel symbols.\\
The linear connection satisfies the properties:
\begin{eqnarray}
{\nabla_{fX + gY}}Z = f {\nabla_{X}}Z + g {\nabla_{Y}}Z \\ \nonumber
{\nabla_{X}}(fY + gZ) = X(f) Y + X(g)Z + f {\nabla_{X}}Y + g {\nabla_{X}}Z
\end{eqnarray}
where $X,Y,Z$ are vectors and $f,g$ are 0-forms (scalar functions) and $X(f) \equiv df(X) \equiv \nabla_{X} f$\\
The covariant exterior derivative is defined as follows:\\
If we have a form $S^{a}$ with one upper index we write:
\be
D S^{a} = dS^{a} + \Lambda^{a}{}_{c} \wedge S^{c}
\ee
with two indices it is written:
\be
D S^{ab} = dS^{ab} + \Lambda^{a}{}_{c} \wedge S^{cb} +  \Lambda^{b}{}_{c} \wedge S^{ac}
\ee
Analogously the case for a generic number of upper indices.\\
If we have one lower index the $D$ operator can be written:
\be
D S_{a} = dS_{a} - \Lambda^{c}{}_{a} \wedge S_{c}
\ee
and with two indices:
\be
D S_{ab} = dS_{ab} - \Lambda^{c}{}_{a} \wedge S_{cb} - \Lambda^{c}{}_{b} \wedge S_{ac}
\ee
Analogously for a general number of lower indices.\\
The case of the mixed tensor is obtained using the latter two particular cases.\\
It is obvious that if we apply $D$ to a form without indices then $D \equiv d$.\\
\\
{\bf Examples}:\\
\\
The formula:
\be
T^{a} = de^{a} + \Lambda^{a}{}_{b} \wedge e^{b}
\ee
can be rewritten as:
\be
T^{a} = De^{a}
\ee
\\
The connection operator applied to a 0-forms is equivalent to the directional derivative $\nabla_{X} f = X(f)$, so we may write since ${\bf g}(Y,Z)$ is a 0-form:
\be
X({\bf g}(Y,Z)) = ({\nabla_{X}}{\bf g})(Y,Z) + {\bf g}({\nabla_{X}}Y,Z) + {\bf g}(Y, {\nabla_{X}}Z)
\ee
If $X \equiv X_{c}, Y \equiv X_{a}, Z \equiv X_{b}$ and defining the non-metricity forms as:
\be
Q_{ab}(X_{c}) = ({\nabla_{X_{c}}}{\bf g})(X_{a}, X_{b})
\ee
we get:
\be
X_{c}(g(X_{a},X_{b})) = Q_{ab}(X_{c}) + \Lambda_{ba}(X_{c}) + \Lambda_{ab}(X_{c}) = dg_{ab}(X_{c})
\ee
since $g(X_{a},X_{b}) = g_{ab}$.\\
Considering the case of the metric theory we get since $Q_{ab} = 0$:
\be
dg_{ab}(X_{c}) = (\Omega_{ab} + \Omega_{ba})(X_{c})
\ee
or:
\be
dg_{ab} = (\Omega_{ab} + \Omega_{ba})
\ee
where $\Omega_{ab}$ is the Riemannian connection.\\
The remaining term gives:
\be
Q_{ab} = -(\lambda_{ab} + \lambda_{ba})
\ee
We can write:
\begin{eqnarray}
Q_{ab} = -(\Lambda_{ab} + \Lambda_{ba}) + (\Omega_{ab} + \Omega_{ba}) = -(\Lambda_{ab} + \Lambda_{ba}) + dg_{ab} = \\ \nonumber
dg_{ab} - \Lambda^{c}{}_{a} \, g_{cb} - \Lambda^{c}{}_{b} \, g_{ac} = Dg_{ab}
\end{eqnarray}
Analogously:
\be
Q^{ab} = - Dg^{ab}
\ee
\\
For the definition and properties of Lie derivatives $\cal{L}$ see for example [44].\\
\\
To conclude we want to present an example of transition from the frame independent approach to the traditional component notation.\\
The stress $3$-form in the electromagnetic theory is given by:
\be
\tau_{a} = \frac{1}{2}[i_{a}F \wedge \star F - i_{a} \star F \wedge F]
\ee
The relation between the stress $3$-forms and the components of the second rank stress tensor is:
\be
T_{ab} = (\star \tau_{a})(X_{b})
\ee
The electromagnetic two forms is written as:
\be
F = \frac{1}{2}F_{ab} \, e^{a} \wedge e^{b}
\ee
Then using the properties of the $\star$ operation, the interior multiplication $i_{a}$ and the wedge product we get the well known expression:
\be
T_{ab} = -[F_{ac}F^{c}{}_{b} + \frac{1}{4}g_{ab}F^{cd}F_{cd}]
\ee 
\newpage
\begin{center}
{\bf Bibliography}
\end{center}
\bigskip
1] R. W. Tucker, C. Wang, \emph{Manifolds: A Maple package for Differential Geometry} (1996).\\
2] J. Scherk and J. H. Schwartz, \emph{Dual Models and the Geometry of Space Time}, Phys. Lett. B {\bf 52B} (1974) 347.\\
3] T. Dereli, R. W. Tucker, \emph{An Einstein-Hilbert action for axi-dilaton gravity in four dimensions}, Class. Quant. Grav. {\bf 12} L31 (1995).\\
4] T. Dereli, M. Onder, R. W. Tucker, \emph{ Solutions for neutral axi-dilaton gravity in four dimensions}, Class. Quant. Grav. {\bf 12} L25 (1995).\\
5] T. Dereli, R. W. Tucker, \emph{Non-metricity induced by dilaton gravity in two dimensions}, Class. Quant. Grav. {\bf 11}, 2575 (1994).\\
6] F. W. Hehl, E. Lord, L. L. Smalley, \emph{Metric-Affine variational principles in general relativity. II Relaxation of the Riemannian constraints}, Gen. Rel. Grav. {\bf 13} (1981) 1037.\\
7] F. W. Hehl, E. W. Mielke: Non-Metricity and Torsion \emph{Proc. 4th Marcel Grossman Meeting on General Relativity, Part A} ed R. Ruffini (Amsterdam: North Holland) (1986) p. 277.\\
8] V. N. Ponomariev, Y. Obukhov, Gen. Rel. Grav. {\bf 14} (1982) 309.\\
9] A. A. Coley, \emph{Analysis of nonmetric theories of gravity. I Electromagnetism}, Phys. Rev. D {\bf 27} (1983) 728.\\
10] A. A. Coley, \emph{Analysis of nonmetric theories of gravity II, the weak equivalence principle}, Phys. Rev. D {\bf 28} (1983) 1829.\\
11] A. A. Coley, \emph{The inconsistency of two nonmetric theories of gravity}, Nuovo Cimento B {\bf 69} (1982) 89.\\
12] M. Gasperini, \emph{A thermal interpretation of the cosmological constant}, Class. Quant. Grav. {\bf 5} (1988) 521.\\
13] J. Stelmach, \emph{Non-metricity driven inflation}, Class. Quant. Grav. {\bf 8} (1991) 897.\\
14] A. K. Aringazin, A. L. Mikhailov, \emph{Matter fields in spacetime with vector non-metricity}, Class. Quant. Grav. {\bf 8} (1991) 1685.\\
15] J. P. Berthias, B. Shabid-Saless, \emph{Torsion and non-metricity in scalar-tensor theories of gravity}, Class. Quant. Grav. {\bf 10} (1993) 1039.\\
16] L. L. Smalley, \emph{Post-Newtonian approximation of the Poincare' gauge theory of gravitation}, Phys. Rev. D {\bf 21} (1980) 328.\\
17] R. W. Tucker, C. Wang, \emph{Dark Matter Gravitational Interaction}, Class. Quant. Grav. {\bf 15} (1998) 933.\\
18] F. W. Hehl, J. D. McCrea, E. W. Mielke, Y. Ne'eman, \emph{Metric-affine gauge theory of gravity: field equations, Noether identities, world spinors, and breaking of dilation invariance}, Phys. Rep. {\bf 258} 1 (1995) .\\
19] N. Yu. Obukhov, V. N Ponomariev, V. V. Zhytnikov, \emph{Quadratic Poincare gauge theory of gravity: a comparison with the general relativity theory}, Gen. Rel. Grav. {\bf 21} (1989) 1107.\\
20] J. F. Pascual-Sanchez, \emph{On the limits of Poincare' gauge theories}, Phys. Lett. A {\bf 108} (1985) 387.\\
21] D. W. Sciama, \emph{The physical structure of General Relativity}, Rev. Mod. Phys. {\bf 36} (1964) 463 and 1103.\\
22] C. H. Wang, \emph{Theory and applications of non-Riemannian Gravitation}, Ph.D. Thesis (Lancaster) (1996).\\
23] R. W. Tucker, C. Wang, \emph{Non-Riemannian gravitational interactions}, Mathematics of Gravitation, Banach Centre Publications, Vol. 41
    Warsawa (1997).\\
24] R. W. Tucker, C. Wang, \emph{Black holes with Weyl charge and non-Riemannian waves}, Class. Quant. Grav. {\bf 12} (1995) 2587.\\
25] T. Dereli, M. Onder, J. Schray, R. W. Tucker and C. H. Wang, \emph{Non-Riemannian gravity and the Einstein-Proca system}, Class. Quant. Grav. {\bf 13} L 103 (1996).\\
26] Yu. N Obukhov, E. J. Vlachynsky, W. Esser and F. W. Hehl, \emph{Effective Einstein theory from metric-affine gravity models via irreducible decompositions}, Phys. Rev. D {\bf 56} 12 (1997) 7769.\\
27] R. Penney, \emph{Generalisation of the Reissner-Nordstrom solution to the Einstein field equations}, Phys. Rev. {\bf 182} 1383 (1969).\\
28] H. Pederson, K. T. Todd, \emph{Three-dimensional Einstein-Weyl geometry}, Adv. in Mathematics {\bf 97} (1993) 74.\\
29] J. D. McCrea, \emph{Irreducible decompositions of non-metricity, torsion, curvature and Bianchi identities in metric-affine spacetimes}, Class. Quant. Grav. {\bf 9} (1992) 553.\\
30] M. A. Melvin, \emph{Pure magnetic and Electric Geons}, Phys. Lett. B 8 1 (1964) 65.\\
31] G. Rosen, \emph{Symmetries of the Einstein-Maxwell Equations}, J. Math. Phys. {\bf 3}, 2 (1962) 313.\\
32] G. W. Gibbons, \emph{Antigravitating Black Hole Solitons with Scalar Hair in n=4 Supergravity}, Nucl. Phys. B {\bf 207} (1982) 337.\\
33] G. W. Gibbons, K. Maeda, \emph{Black Holes and Membranes in Higher-Dimensional theories with Dilaton fields}, Nucl. Phys. B {\bf 298} (1988) 741.\\
34] D. Garfinkel, G. T. Horowitz, A. Strominger, \emph{Charged black holes in string theory}, Phys. Rev. D {\bf 43} 10 3140 (1991).\\
35] M. A. Vandyck, \emph{Maxwell's equations in spaces with non-metricity and torsion}, J. Phys. A: Math. Gen. {\bf 29} 2245 (1996).\\
36] R. A. Puntingam, F. W. Hehl, \emph{Maxwell's theory on a post-Riemannian spacetime and the equivalence principle}, Class. Quant. Grav. {\bf 14} (1997) 1347.\\
37] F. Avignone, \emph{Laboratory limits on solar axions}, Phys. Rev. D {\bf 35} (1987) 2752.\\
38] M. S. Turner, \emph{Early universe thermal production of not-so-invisible axions}, Phys. Rev. Lett. {\bf 59} (1987) 2489.\\
39] E.Kolb, M.Turner,\emph{The early Universe}(1990) Addison Wesley Publishing.\\
40] N. Banerjee and S. Sen, \emph{No Scalar hair theorem for a charged spherical black hole}, Phys. Rev. D {\bf 58} 1 104024 (1998).\\
41] C. Brans, R. H. Dicke, \emph{Mach's principle and a relativistic theory of gravitation}, Phys. Rev. {\bf 124} 925 (1961).\\
42] J. D. Bekenstein, \emph{Exact solutions of Einstein-Conformal Scalar Equations}, Ann. Phys. {\bf 82} (1974) 535.\\
43] R. Adler, M. Bazin, M. Schiffer, \emph{Introduction to General Relativity} 1975 (New York: McGraw-Hill).\\
44] See for example, I. Benn, R. W. Tucker, \emph{An introduction to spinors and geometry}, (1987), Adam Hilger.\\
45] E. Kroner, \emph{Continuum theory of defects} in Physics of Defects, Les Houches, Session XXXV, 1980, R. Balian et al. eds. (North-Holland, Amsterdam 1981) 215.\\
46] E. Kroner, \emph{The role of differential geometry in the mechanics of solids}, in Proc. 5th Nat. Congr. Theor. Appl. Mech. Vol.1, (Bulgarian Acad. Sci. Sofia 1985) 352-362. Since this reference is not easily reachable, we quote from the abstract:\\ 
Both the statical and the geometrical field equations of nonlinear elastostatics are conveniently formulated in terms of differential geometry ...... Bravais crystals are described by the general affine differential geometry such that point defects correspond to the nonmetric part of the connection, line defects are represented by the torsion.\\
47] Y. Neeman, D. Sijacki, \emph{Unified affine gauge theory of gravity and strong interactions with finite and infinite $GL(4,R)$ spinor fields}, Ann. Phys. {\bf 120} (1979) 292.\\
48] Y. Neeman, D. Sijacki, \emph{Hadrons in an $SL(4,R)$ classification}, Phys. Rev. D {\bf 37} (1988) 3267.\\
49] P. J. Steinhart, \emph{Recent advances in extended inflationary cosmology}, Class. Quant. Grav. {\bf 10} (1993) L33.\\
50] H. Weyl, Geometrie und Physik, Naturwissenschaften {\bf 19} (1931) 49.\\
51] R. Utiyama, \emph{Invariant theoretical interpretation of interaction}, Phys. Rev. {\bf 101} (1956) 1597.\\
52] T. W. B. Kibble, \emph{Lorentz invariance and the gravitational field}, J. Math. Phys. {\bf 2} (1961) 212.
\end{document}